%% file: main.tex
\RequirePackage{fix-cm}

\documentclass{article}
\newcommand{\qed}{\hfill$\blacksquare$}
\usepackage{graphicx}
\usepackage{tikz-qtree}
\usepackage{tikz}
\usetikzlibrary{bayesnet}
\usepackage{xfrac}
\usepackage{subcaption}
\usepackage[colorlinks=true,citecolor=blue]{hyperref}  
\input{defs}
\begin{document}

\title{Causal Structure Learning:\\ a Combinatorial Perspective
}


\author{
Chandler Squires         
\and
Caroline Uhler
}



\maketitle

\begin{abstract}
In this review, we discuss approaches for learning causal structure from data, also called \textit{causal discovery}.
In particular, we focus on approaches for learning directed acyclic graphs (DAGs) and various generalizations which allow for some variables to be unobserved in the available data.
We devote special attention to two fundamental combinatorial aspects of causal structure learning.
First, we discuss the structure of the search space over causal graphs.
Second, we discuss the structure of \textit{equivalence classes} over causal graphs, i.e., sets of graphs which represent what can be learned from observational data alone, and how these equivalence classes can be refined by adding \textit{interventional} data.
%
\end{abstract}

\section{Introduction}\label{sec:intro}

Many important scientific, sociological, and technological questions are fundamentally \textit{causal}: ``which genes regulate one another?", ``how would raising minimum wage affect unemployment rate?", ``which treatment most effectively prolongs the lifespan of breast cancer patients?".
In each case, answering the question requires predicting how a system, e.g., a cell, economy, or human body, will react to external manipulation.
\textit{Structural causal models} can be used to formalize such questions, to create algorithms that determine whether such questions can be answered from available data sources, and to develop general-purpose methods for learning the answers to such questions.
In the framework of structural causal models, a \textit{directed graph} is used to reflect how the variables in these models depend causally on one another.
Each node $i$ of the directed graph is associated with a variable $X_i$, and an edge $i \to j$ indicates that the variable $X_i$ is a direct cause of the variable $X_j$.
In some special, well-studied settings, background knowledge and human reasoning can be used to propose plausible directed graph models.
However, in large systems such as gene regulatory networks, the directed graph is \textit{not} known a priori, making it necessary to develop methods for \textit{learning} the graph from data.
Once this graph is learned, it can be used to predict the effects of interventions or distributional shifts, in contrast to traditional machine learning methods which can only make predictions on inputs that come from the same distribution as the training data.

The problem of learning such a causal graph from data, known as \textit{causal structure learning} (or \textit{causal discovery}), has been the focus of much recent work in computer science, statistics, and bioinformatics
, covered in a number of recent reviews \cite{daly2011learning,drton2017structure,meinshausen2016methods,heinze2018causal,glymour2019review}.
Compared to these reviews, we here emphasize the \textit{combinatorial} aspects of causal structure learning, including characterizations of equivalence classes of graphs, computing the size and number of these equivalence classes, and how the characterization and properties are influenced by the presence of latent variables or interventional data.
After discussing these topics, we will cover methods for causal structure learning which are based heavily on the combinatorial structure over the space of directed graphs.
Focusing on this combinatorial structure has three significant advantages:
\begin{enumerate}
    \item Causal structure learning can be dramatically simplified when fixing some combinatorial aspect of the problem, such as the ordering of the variables.
    
    \item Understanding the combinatorial aspects of structure learning allows a number of different methods to be synthesized into a single framework, and eases future methodological development.
    
    \item Insights into the combinatorial aspects of structure learning are also useful for other tasks, such as experimental design.
\end{enumerate}
The framework provided by the combinatorial viewpoint encompasses methods for learning causal models with unobserved variables, as well as methods for learning from a combination of observational and interventional data.
The second point is especially important, since interventional data is often crucial for identifying the true causal model, and subsequently using the causal model for predicting the effects of interventions or distributional shifts.

\section{Structural Causal Models}\label{sec:preliminaries}

A structural causal model 
defines causal relationships over a set of random variables $\{ X_i \}_{i=1}^p$.
These relationships are summarized by a directed acyclic graph (DAG) $\cG$ over nodes $i = 1, \ldots, p$, where the node $i$ in $\cG$ is associated with the variable $X_i$.
Given a DAG $\cG$, we let $\pa_\cG(i)$ denote the \textit{parents} of the node $i$, i.e., $\pa_\cG(i) = \{ j \mid j \to i ~\textrm{in}~\cG \}$.
Then, a (Markovian) \textit{structural causal model} (SCM) \cite{peters2017elements} with \textit{causal graph} $\cG$ consists of a set of \textit{endogenous variables} $\{ X_i \}_{i=1}^p$, a set of \textit{exogenous variables} $\{ \epsilon_i \}_{i=1}^p$, a product distribution $\bbP_\epsilon$ over the exogenous variables, and a set of \textit{structural assignments} $\{ f_i \}_{i=1}^p$.
In particular, the structural assignment $f_i$ asserts the relation $X_i = f_i(X_{\pa_\cG(i)}, \epsilon_i)$.
Via these structural assignments, the distribution $\bbP_\epsilon$ over the exogenous variables induces a distribution $\bbP_X$ over the endogenous variables, called the \textit{entailed distribution} \cite{peters2017elements}.
In particular, we have $\bbP_X(X_i \mid X_{\pa_\cG(i)}) = \bbE_{\epsilon_i} [\kron_{X_i = f_i(X_{\pa_\cG(i)}, \epsilon_i)} \mid X_{\pa_\cG(i)}]$ and
\begin{equation}\label{eq:factorization}
	\bbP_X(X) = \prod_{i=1}^p \bbP_X(X_i \mid X_{\pa_\cG(i)}).
\end{equation}

\begin{example}[A simple structural causal model of genetic inheritance]\label{ex:mouse-scm}
As a running example, we will consider a simplified model of genetic inheritance of weight among a family of mice.
Let $X_2$ and $X_3$ represent the weights, in grams, of an unrelated male and female mouse, respectively.
Let $X_4$ represent the weight of their offspring, and $X_5$ represent the weight of the offspring's offspring.
Finally, let $X_1$ be a binary variable representing whether the two parent mice are genetically modified for increased weight.
Assume that these variables are related via the following set of assignments:

\begin{align*}
    X_1 &= \epsilon_1
    &&\epsilon_1 \sim \Ber(0.5)
    \\
    X_2 &= \epsilon_2 + 2 X_1
    &&\epsilon_2 \sim \cN(25, 1)
    \\
    X_3 &= \epsilon_3 + 2 X_1
    &&\epsilon_2 \sim \cN(20, 1)
    \\
    X_4 &= \sfrac{1}{2} \left( X_2 + X_3 \right) + \epsilon_4
    &&\epsilon_4 \sim \cN(0, 1)
    \\
    X_5 &= X_4 + \epsilon_5
    &&\epsilon_5 \sim \cN(0, 2)
\end{align*}
where the set of $\epsilon$ are mutually independent.
The parent sets are $\pa_\cG(1) = \varnothing, \pa_\cG(2) = \{ 1 \}$ $\pa_\cG(3) = \{ 1 \}$, $\pa_\cG(4) = \{ 2, 3 \}$, and $\pa_\cG(5) = \{ 4 \}$.
The causal graph is given in \rref{fig:scm-and-minimal-imap}, and 
\begin{align*}
    \bbP_X(X) = 
    \Ber(X_1; 0.5) 
    &\times \cN(X_2; 25 + 2 X_1, 1) 
    \times \cN(X_3; 20 + 2 X_1, 1) 
    \\
    &\times \cN(X_4; \sfrac{1}{2} (X_2 + X_3), 1) 
    \times \cN(X_5; X_4, 2) 
\end{align*}
is the entailed distribution. \qed
\end{example}

The above definition of structural causal models can be generalized in at least two ways.
First, one may remove the assumption that the distribution over the exogenous variables is a product distribution, i.e., one may allow dependence between $\epsilon_i$ and $\epsilon_j$ for $i \neq j$.
Such SCMs are called \textit{semi-Markovian}, and are taken as the basic definition of SCMs by some authors \cite{pearl2009causality}.
Instead of allowing for dependencies between exogenous variables, we use Markovian SCMs as the basic definition, and assume that any unmodeled dependence between endogenous variables is due to some other \textit{unobserved} endogenous variables, which we will cover in \rref{subsec:graphical-unobserved}.
Second, one may remove the assumption that $\cG$ is acyclic.
The assumption of acyclicity is natural when considering endogenous variables which are defined at certain time points, since the intuitive notion of causality dictates that a cause precedes any of its effects.
However, if the endogenous variables are not well-defined in time, e.g., if they represent the average state of a system in equilibrium, then feedback loops may occur.
We will briefly discuss recent progress on causal structure learning for cyclic causal models in \rref{sec:discussion}.
%

\subsection{Markov properties and Markov equivalence in DAGs}

Given a DAG $\cG$, the set of distributions $\bbP_X$ that factorize according to \rref{eq:factorization} are said to follow the \textit{Markov factorization property} with respect to $\cG$.
Depending on assumptions on the structural equations $\{ f_i \}_{i=1}^p$ and the exogenous variables $\{ \epsilon_i \}_{i=1}^p$, the Markov factorization property implies many other testable properties of the distribution $\bbP_X$.
For instance, the \textit{entire} set of conditional independence statements entailed by the Markov factorization property can be characterized simply in terms of a graphical criterion, known as \textit{d-separation}, that can be read off from the DAG $\cG$.
The definition of d-separation relies on the notion of a \textit{collider} along a path from $i$ to $j$.
Given a path $\gamma = \langle \gamma_1 = i, \gamma_2, \ldots, \gamma_M = j \rangle$ from $i$ to $j$, the node $\gamma_m$ is a \textit{collider} if $\gamma_{m-1} \to \gamma_m \leftarrow \gamma_{m+1}$, i.e., two arrowheads ``collide" at $\gamma_m$.
Then, a path $\gamma$ \textit{d-connects} $i$ and $j$ given the set $C \subseteq [p] \setminus \{ i, j \}$ if:
\begin{enumerate}
    \item All non-colliders on the path do not belong to $C$.
    
    \item All colliders on the path either belong to $C$, or have a descendant which belongs to $C$.
\end{enumerate}
Finally, $i$ and $j$ are \textit{d-connecting} given $C$ if there exists any d-connecting path given $C$; otherwise, they are \textit{d-separated}.
We denote that $i$ and $j$ are d-separated in $\cG$ given $C$ via $i \indep_\cG j \mid C$.
We denote the complete set of d-separation statements in a DAG $\cG$ as $\indepmodel(\cG)$; i.e.,
\begin{equation*}
    \indepmodel(\cG) = \{ (i, j, C) \mid i, j \in [p], C \subseteq [p] \setminus \{i, j\}, i \indep_\cG j \mid C \}.
\end{equation*}

\begin{figure}
    \centering
    \includegraphics[width=\textwidth]{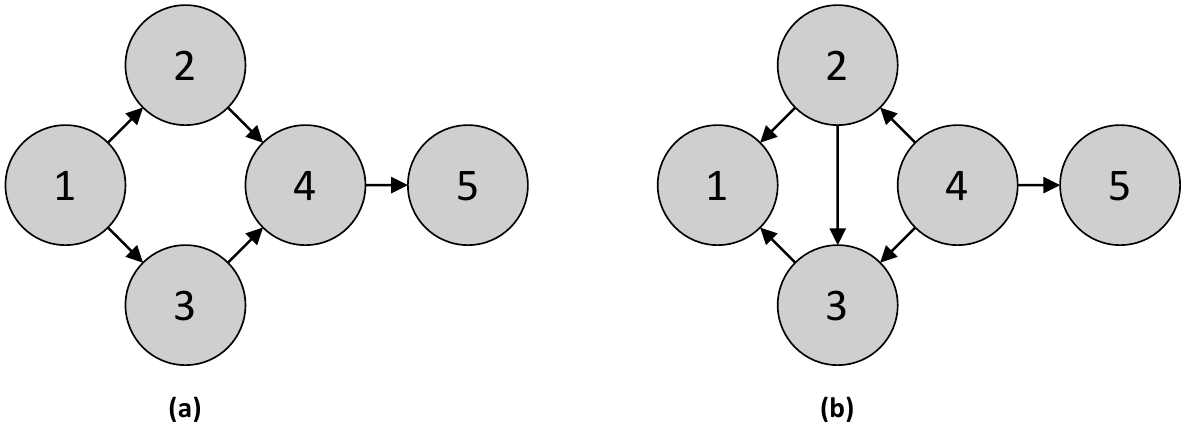}
    \caption{
    \textbf{(a)} The causal graph $\cG^*$ for the structural causal model in \rref{ex:mouse-scm}.
    \textbf{(b)} A minimal I-MAP $\cG_2$ for $\cG^*$, described in \rref{ex:multiple-minimal-imaps}.
    }
    \label{fig:scm-and-minimal-imap}
\end{figure}

\begin{example}[d-connection and d-separation]\label{ex:d-connection}
In $\cG^*$ from \rref{fig:scm-and-minimal-imap}(a), there are two paths between 2 and 3, the path $\gamma_1 = 2 \leftarrow 1 \rightarrow 3$, and the path $\gamma_2 = 2 \to 4 \leftarrow 3$.
For $C = \varnothing$, $\gamma_1$ is a d-connecting path between 2 and 3, since 1 is a non-collider and does not belong to $C$, while $\gamma_2$ is \textit{not} a d-connecting path, since 4 is a collider but neither 4 nor 5 is in $C$.
Thus, 2 and 3 are d-connected given $C = \varnothing$.
For $C = \{ 1 \}$, neither $\gamma_1$ nor $\gamma_2$ are d-connecting paths, so 2 and 3 are d-separated given $C = \{ 1 \}$.
Finally, for any $C$ containing 4 or 5, $\gamma_2$ is a d-connecting path between 2 and 3.
Thus, 2 and 3 are d-connected given $C = \{ 4 \}$, $C = \{ 5 \}$, $C = \{ 1, 4 \}$, etc. \qed
\end{example}

	Given a distribution $\bbP_X$, we call $X_i$ and $X_j$ \textit{conditionally independent} given $X_C$ if $\bbP_X(X_i, X_j \mid X_C) = \bbP_X(X_i \mid X_C) \bbP_X(X_j \mid X_C)$.
	This is denoted by $i \indep_{\bbP_X} j \mid C$.
	We denote the set of all conditional independence statements in $\bbP_X$ as
	\begin{equation*}
		\indepmodel(\bbP_X) = \{ (i, j, C) \mid i, j \in [p], C \subseteq [p] \setminus \{i, j\}, i \indep_{\bbP_X} j \mid C \}.
	\end{equation*}
If all d-separation statements in the DAG $\cG$ hold as conditional independence statements in $\bbP_X$,  i.e., $\indepmodel(\cG) \subseteq \indepmodel(\bbP_X)$, then $\bbP_X$ is said to satisfy the \textit{global Markov property} with respect to $\cG$.
Suppose that $\bbP_X$ has a density with respect to some product measure.
Then, without any additional assumptions on the structural equations or the distributions of exogenous variables, the Markov factorization property and the global Markov property are equivalent \cite{maathuis2018handbook}.

Conversely, a given distribution $\bbP_X$ may satisfy the global Markov property with respect to many different DAGs.
These DAGs are called \textit{independence maps (I-MAPs)} of the distribution $\bbP_X$.
As an extreme example, the complete graph implies \textit{no} conditional independencies in $\bbP_X$, so it is an I-MAP of all distributions.
However, the complete graph does not capture any of the independence structure in $\bbP_X$.
For a variety of purposes, including computational and statistical efficiency in inference and estimation, it is preferable to find a DAG $\cG$ that captures as many of the independences of $\bbP_X$ as possible.
This intuition is captured in the definition of a \textit{minimal} I-MAP for $\bbP_X$, which is an I-MAP $\cG$ of $\bbP_X$, such that the deletion of \textit{any} edge will result in a new DAG $\cG'$ which is no longer an I-MAP for $\bbP_X$.
The following example shows that a distribution $\bbP_X$ can have several minimal I-MAPs.

\begin{example}[A distribution $\bbP_X$ can have multiple minimal I-MAPs]\label{ex:multiple-minimal-imaps}
Let $\bbP_X$ be the distribution in \rref{ex:mouse-scm}.
Then the DAG $\cG^*$ in \rref{fig:scm-and-minimal-imap}(a) is a minimal I-MAP for $\bbP_X$.
To see this, we consider the deletion of each edge.
Deleting $1 \to 2$ or $1 \to 3$ implies that $X_1 \indep X_2$, or $X_1 \indep X_3$, respectively, both of which are false.
Similarly, deleting $2 \to 4$ or $3 \to 4$ implies that $X_2 \indep X_4$, or $X_3 \indep X_4$, respectively, but both are false.
Finally, deleting $4 \to 5$ implies that $X_4 \indep X_5$, which is again false.

$\bbP_X$ has other minimal I-MAPs, including the DAG $\cG_2$ in \rref{fig:scm-and-minimal-imap}(b).
Deleting $2 \to 1$ and $3 \to 1$ implies $X_2 \indep X_1 \mid X_4, X_3$ and $X_3 \indep X_1 \mid X_2$, respectively, both of which are false.
Deleting $2 \to 3$ implies that $X_2 \indep X_3 \mid X_4$, deleting $4 \to 2$ implies $X_4 \indep X_2$, deleting $4 \to 3$ implies $X_4 \indep X_3 \mid X_2$, and deleting $4 \to 5$ implies $X_4 \indep X_5$, showing that $\cG_2$ is indeed minimal. \qed
\end{example}


Suppose $\bbP_X$ is entailed by an SCM with causal graph $\cG^*$.
Since $\bbP_X$ may have multiple minimal I-MAPs, it is natural to ask, under some set of assumptions, whether $\cG^*$ can be distinguished from the other minimal I-MAPs, and if not, whether a small \textit{subset} of the minimal I-MAPs can be distinguished as candidates for $\cG^*$.
As we will discuss in \rref{sec:algorithms}, without assumptions on the functional forms of the structural assignments $f_i$, one cannot in general distinguish $\cG^*$ from all other graphs using only $\bbP_X$.
In particular, two DAGs $\cG$ and $\cG'$ with the same set of d-separation statements (i.e., $\indepmodel(\cG) = \indepmodel(\cG')$) are called \textit{Markov equivalent}, and we denote this by $\cG \approx_\cM \cG'$.
The set of all DAGs that are Markov equivalent to $\cG^*$ is called the \textit{Markov equivalence class} (\emph{MEC}) of $\cG^*$, denoted $\cM(\cG^*)$, and $\bbP_X$ can in general only identify $\cG^*$ up to $\cM(\cG^*)$.

\begin{figure}
    \centering
    \includegraphics[width=\textwidth]{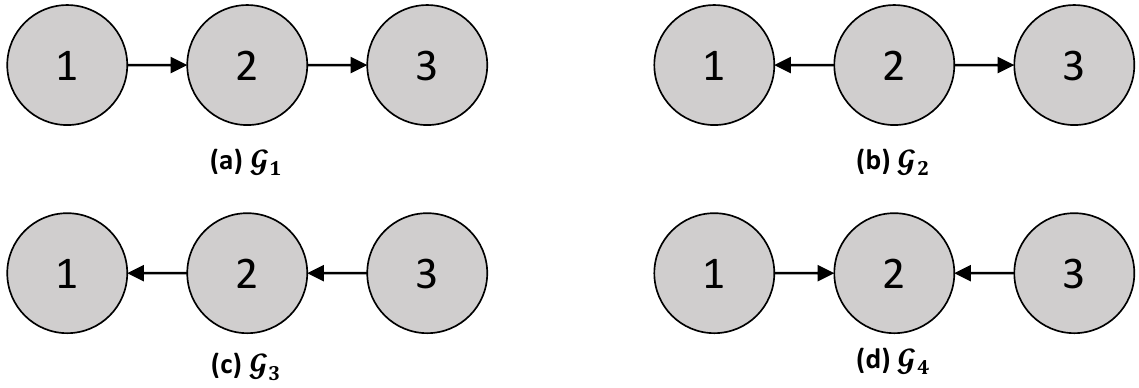}
    \caption{\textbf{(a), (b), (c)} Three Markov equivalent graphs from \rref{ex:markov-equivalence}. \textbf{(d)} A fourth graph that is not Markov equivalent to the other three.}
    \label{fig:markov-equivalence}
\end{figure}

\begin{example}[Markov equivalence]\label{ex:markov-equivalence}
The three DAGs in \rref{fig:markov-equivalence}(a,b,c) are all Markov equivalent to one another, since for all three graphs, the only d-separation statement is that $1$ and $3$ are d-separated given $2$.
However, the DAG in \rref{fig:markov-equivalence}(d) is \textit{not} a member of the same MEC, since in $\cG_4$, 1 and 3 are (unconditionally) d-separated, but are d-connected given 2. \qed
\end{example}

However, under certain assumptions, it is possible to distinguish the set $\cM(\cG^*)$ from all other minimal I-MAPs of $\bbP_X$.
This is the case under the \textit{sparsest Markov representation (SMR)} assumption \cite{raskutti2018learning}, which states that, for any minimal I-MAP $\cG'$ of $\bbP_X$ such that $\cG' \not\in \cM(\cG^*)$, we have $|\cG'| > |\cG^*|$, where $|\cG|$ denotes the number of edges in $\cG$.
Under this assumption, $\cM(\cG^*)$ can be identified by enumerating over minimal I-MAPs of $\bbP_X$ and picking the sparsest minimal I-MAP.

More generally, to identify $\cM(\cG^*)$, structure learning algorithms require some form of \textit{faithfulness} assumption.
The strongest such assumption, referred to simply as \textit{the faithfulness assumption}, is exactly the converse to the global Markov property: all conditional independence statements in $\bbP_X$ must hold as d-separation statements in $\cG^*$, i.e., $\indepmodel(\bbP_X) = \indepmodel(\cG^*)$.
The faithfulness assumption is a ``genericity" assumption in the sense that for parametric models, such as linear Gaussian models, the set of parameters which violate the faithfulness assumption is of Lebesgue measure zero \cite{spirtes2000causation}.
This is demonstrated by the following example.
\begin{example}
Consider the distribution $\bbP_X$ entailed by the following SCM:
\begin{align*}
    X_1 &= \epsilon_1
    &&\epsilon_1 \sim \cN(0, 1)
    \\
    X_2 &= \epsilon_2 + \beta_{12} X_1
    &&\epsilon_2 \sim \cN(0, 1)
    \\
    X_3 &= \epsilon_3 + \beta_{13} X_1
    &&\epsilon_2 \sim \cN(0, 1)
    \\
    X_4 &= \beta_{24} X_2 + \beta_{34} X_3 + \epsilon_4
    &&\epsilon_4 \sim \cN(0, 1)
\end{align*}
Denoting the corresponding causal graph by $\cG$, then the d-separation statements are given by $\indepmodel(\cG) = \{ (1, 4, \{ 2, 3 \}), (2, 3, \{ 1 \}) \}$.
However, if $\beta_{12} \beta_{24} + \beta_{13} \beta_{34} = 0$, then $\Cov(X_1, X_2) = 0$, so by Gaussianity, we have that $1 \indep_{\bbP_X} 4$, i.e., $(1, 4, \varnothing) \in \indepmodel(\bbP_X)$ but $(1, 4, \varnothing) \not\in \indepmodel(\cG^*)$.
The set of parameters $(\beta_{12}, \beta_{13}, \beta_{24}, \beta_{34})$ satisfying this equality is of Lebesgue measure zero. \qed
\end{example}

In this example, the effect of $X_1$ on $X_4$ along the paths $1 \to 2 \to 4$ and $1 \to 3 \to 4$ perfectly ``cancels out".
While perfect cancellation may only occur for very specific parameters, structure learning algorithms do not have direct access to $\bbP_X$, and must \textit{test} for conditional independence using samples from $\bbP_X$.
Thus, \textit{near} cancellations, e.g., if $\beta_{12}\beta_{24} + \beta_{13} \beta_{34} = 0.0015$, may be indistinguishable from cancellations at small sample sizes.
To overcome noise and provide finite-sample or high-dimensional guarantees for structure learning algorithms, it is necessary to make stronger assumption, such as \textit{strong} faithfulness \cite{zhang2002strong}, which assumes that the (conditional) mutual information between d-connected variables is bounded away from zero.
However, the set of parameters which violate the strong faithfulness assumption can have large Lebesgue measure \cite{uhler2013geometry}.
This has motivated the development of structure learning algorithms under assumptions that only require some \textit{subset} of the missing d-separation statements in $\indepmodel(\cG)$ to hold ``strongly" in $\bbP_X$, thus reducing the size of the set of violating parameters.
Such assumptions, including e.g.~a strong version of the SMR assumption, are reviewed and compared in \cite{zhang2016three,raskutti2018learning}.

%
Since in general $\cG^*$ can only be identified up to its MEC, the natural search space for causal structure learning algorithms is over MECs, rather than DAGs.
Consequently, characterizing the structure \textit{within} and \textit{between} MECs has been an important problem for developing structure learning algorithms.
We will discuss useful characterizations of the MEC in \rref{sec:identifiability}.
One way to overcome the limitations on learning from observational data is by using data from \textit{interventions}, which we now formalize.

\subsection{Interventions and Interventional Markov Equivalence}\label{subsec:interventions}
To formalize the effect of an \textit{intervention} $I$ in an SCM, we consider a new \textit{interventional SCM} where we modify some subset of the structural assignments and/or the distributions of exogenous noise variables, without introducing new nodes into any of the parent sets.
If a node $i$ has either its structural assignment $f_i$ or the distribution of its exogenous noise $\epsilon_i$ modified by intervention $I$, it is called a \textit{target} of the intervention, and we write $i \in I$.
The new SCM induces a different distribution $\bbP_X^I$ on $X$, called the \textit{interventional distribution}, which takes the form
\begin{equation}\label{eq:interventional-factorization}
    \bbP_X^I(X) 
    = 
    \prod_{i \not\in I} \bbP_X(X_i \mid X_{\pa_\cG(i)})
    \prod_{i \in I} \bbP_X^I(X_i \mid X_{\pa_\cG(i)}).
\end{equation}

In general, an intervention consists of any modification of the structural assignment or exogenous noise.
To distinguish this most general form of intervention from more stringent definitions of intervention, we will follow \cite{peters2017elements} and call these \textit{soft} interventions (also referred to as \textit{mechanism changes} in \cite{tian2001causal}).
Particular subclasses of interventions have generated special interest.
Most significantly, a \textit{hard} intervention, also called a \textit{perfect}, \textit{surgical} \cite{campbell2007interventionist}, or \textit{structural} \cite{eberhardt2007interventions} intervention, is one which completely removes the dependence of a target $X_i$ on its parents.
However, perfect interventions allow for the target to depend on $\epsilon_i$, so that the target's value may still be random, i.e., the interventional distribution is
\begin{equation}\label{eq:perfect-interventional-factorization}
    \bbP_X^I(X) 
    = 
    \prod_{i \not\in I} \bbP_X(X_i \mid X_{\pa_\cG(i)})
    \prod_{i \in I} \bbP_X^I(X_i).
\end{equation}

More extremely, if the structural assignment of $X_i$ is changed to a constant $a_i$, then there is no randomness left in $X_i$. 
Such a perfect intervention is called a \textit{do-intervention} \cite{meinshausen2016methods}.
In this case, the interventional distribution is
\begin{equation}\label{eq:do-interventional-factorization}
    \bbP_X^I(X) 
    = 
    \prod_{i \not\in I} \bbP_X(X_i \mid X_{\pa_\cG(i)})
    \prod_{i \in I} \kron_{X_i = a_i}.
\end{equation}

\begin{example}[The interventional SCM for mouse genetic modification]\label{ex:mouse-intervention}
Suppose we implement an intervention on the model in \rref{ex:mouse-scm}, where we edit the genome of the offspring mouse to reduce its weight.
In particular, the effect of this intervention is to change the distribution of $\epsilon_4$ to $\cN(-10, 0.1)$.
The interventional distribution is
\begin{align*}
    \bbP_X(X) = 
    \Ber(X_1; .5) 
    &\times \cN(X_2; 25 + 2 X_1, 1) 
    \times \cN(X_3; 20 + 2 X_1, 1) 
    \\
    &\times \cN(X_4; \sfrac{1}{2} (X_2 + X_3) - 10, 1) 
    \times \cN(X_5; X_4, 1).
\end{align*}

This intervention is \textit{not} a perfect intervention, since $X_4$ still depends on its parent $X_2$ and $X_3$.
If instead the genetic modification perfectly ensures that the offspring weights 15 grams, i.e. $X_4 = 15$ always, then the intervention would be a perfect intervention - in particular, a \textit{do}-intervention.
In this case, the interventional distribution becomes
\begin{align*}
    \bbP_X(X) = 
    \Ber(X_1; .5) 
    &\times \cN(X_2; 25 + 2 X_1, 1) 
    \times \cN(X_3; 20 + 2 X_1, 1) 
    \\
    &\times \kron_{X_4 = 15}
    \times \cN(X_5; X_4, 1),
\end{align*}
where $X_4$ does not depend on its parents anymore. \qed
\end{example}

The causal DAG also implies relationships between the observational and interventional distributions.
The simplest approach to deriving these relationships is to \textit{extend} the DAG to include variables which represent different interventions, as proposed in \cite{yang2018characterizing} and used by \cite{squires2020permutation}.
This approach can be seen as an important special case of the \textit{Joint Causal Inference (JCI)} framework \cite{mooij2020joint}.
For a single intervention $I$ with targets $T$, this can be achieved by adding a node $\zeta$ with children $T$. $\zeta$ represents a binary variable, where $\zeta = 1$ denotes that a sample comes from the intervention $I$, and $\zeta = 0$ denotes otherwise.

\begin{figure}
    \centering
    \includegraphics[width=\textwidth]{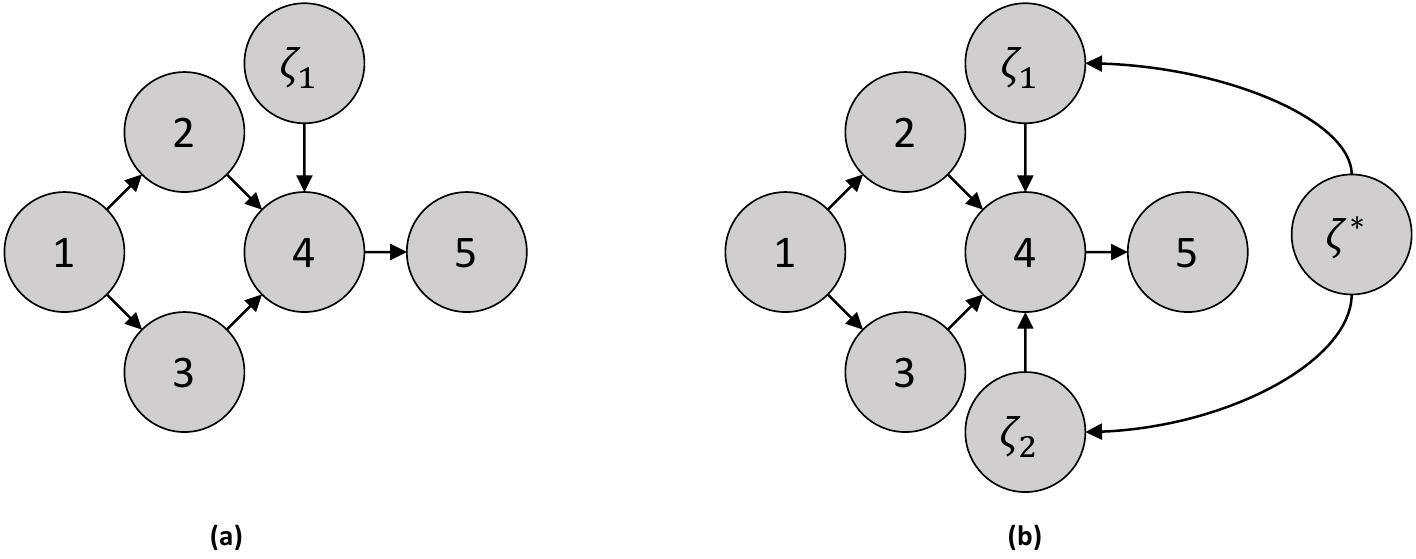}
    \caption{
    \textbf{(a)} The $\cI$-DAG from \rref{ex:binary-encode-intervention}.
    \textbf{(b)} The $\cI$-DAG from \rref{ex:binary-encode-intervention-set}.
    }
    \label{fig:interventional-dags}
\end{figure}

\begin{example}[Binary encoding of an intervention]\label{ex:binary-encode-intervention}
Consider the intervention $I_1$ in \rref{ex:mouse-intervention}, where the intervention is applied with probability $0.5$.
Then the joint distribution over $X$ and $\zeta$ is
\begin{align*}
    \bbP_{X,\zeta}(X,\zeta) = 
    \Ber(\zeta; .5)
    &\times \Ber(X_1; .5) 
    \times \cN(X_2; 25 + 2 X_1, 1) 
    \times \cN(X_3; 20 + 2 X_1, 1) 
    \\
    &\times \cN(X_4; \sfrac{1}{2} (X_2 + X_3) - 10 \zeta, 1) 
    \times \cN(X_5; X_4, 1) .
\end{align*}
The causal DAG for $\zeta, X_1, X_2, X_3, X_4, X_5$ is shown in \rref{fig:interventional-dags}(a).
The node 5 is d-separated from $\zeta$ given 4.
Therefore, $\bbP(X_5 \mid X_4, \zeta = 1) = \bbP(X_5 \mid X_4, \zeta = 0)$, i.e., $\bbP_X(X_5 \mid X_4) = \bbP_X^{I_1}(X_5 \mid X_4)$. \qed
\end{example}

To generalize to \textit{multiple} interventions, we add a node for each intervention.
In particular, consider a set of interventions $\cI = \{ I_1, \ldots, I_M \}$.
For the intervention $I_m$ with targets $T_m$, we introduce a node $\zeta_m$ with children $T_m$.
Again, $\zeta_m = 1$ denotes that the sample comes from the intervention $I_m$, and $\zeta_m = 0$ otherwise.
However, each sample can only be generated from a single intervention, i.e., $\zeta_m = 1$ for at most one $m$.
To reflect this constraint, we include a final node $\zeta^*$, which takes values in $0, 1, \ldots, M$, to indicate which intervention the sample comes from, i.e., $\zeta_m = 1$ if and only if $\zeta^* = m$.
Thus, if $\zeta^* = 0$, the sample comes from the observational distribution.
The resulting DAG is called the \textit{interventional DAG} ($\cI$-DAG) \cite{yang2018characterizing}.

\begin{example}[Binary encoding of a set of interventions]\label{ex:binary-encode-intervention-set}
Let $I_1$ be the intervention in \rref{ex:mouse-intervention}, and let $I_2$ be an intervention which changes the distribution of $\epsilon_4$ to $\cN(-5, 0.1)$.
Suppose each intervention has a 40\% chance of being applied. 
In the remaining 20\% of the time, no intervention takes place.
Then the joint distribution over $X, \zeta_1, \zeta_2$, and $\zeta^*$ is
\begin{align*}
    \bbP_{X,\zeta}(X,\zeta) = 
    &\Cat(\zeta^*; (0, 1, 2), (0.2, 0.4, 0.4))
    \times \left( 1 - \kron_{\zeta_1 = 1, \zeta^* \neq 1} \right)
    \times \left( 1 - \kron_{\zeta_2 = 1, \zeta^* \neq 2} \right)
    \\
    \times &\Ber(X_1; .5) 
    \times \cN(X_2; 25 + 2 X_1, 1) 
    \times \cN(X_3; 20 + 2 X_1, 1) 
    \\
    \times &\cN(X_4; \sfrac{1}{2} (X_2 + X_3) - 10 \zeta_1 - 5 \zeta_2, 1) 
    \times \cN(X_5; X_4, 1).
\end{align*}
The causal DAG for $\zeta_1, \zeta_2, \zeta^*, X_1, X_2, X_3, X_4, X_5$ is shown in \rref{fig:interventional-dags}(b). \qed
\end{example}

Following \cite{yang2018characterizing} we define a \textit{conditional invariance} statement to be a conditional independence statement where the conditioning set includes intervention variables, e.g., $\bbP_{X,\xi}(X_i \mid X_C, \xi^* = m) = \bbP_{X,\xi}(X_i \mid X_C, \xi^* = 0)$.
This statements posits that a conditional distribution in the $m$-th interventional setting is the same as it is in the observational setting, i.e., the conditional distribution is \textit{invariant} under the intervention.
A set of observational and interventional distributions satisfies the \textit{$\cI$-Markov property} with respect to a DAG $\cG$ and a set of interventions $\cI$ if it satisfies the global Markov property with respect to $\cG$, and satisfies all conditional invariance statements entailed by the $\cI$-DAG.
Similarly to the observational case, given a set $\cI$ of interventions, if two DAGs $\cG$ and $\cG'$ entail the same set of conditional independence and conditional invariance statements, we call them \textit{$\cI$-Markov equivalent}, denoted $\cG \approx_{\cM_\cI} \cG'$.
The resulting $\cI$-Markov equivalence class ($\cI$-MEC) is thus a (not necessarily strict) subset of the MEC, as demonstrated by the following example.
\begin{example}[Interventional Markov equivalence]
Given the intervention set $\cI = \{ I_1 \}$ for $I_1$ with target 1, the graphs $\cG_2$ and $\cG_3$ in \rref{fig:markov-equivalence} are $\cI$-Markov equivalent, since they both entail the invariance statements $\bbP^{I_1}(X_2) = \bbP(X_2)$ and $\bbP^{I_1}(X_3) = \bbP(X_3)$.
However, $\cG_1$ does \textit{not} entail these invariance statements, so it is not $\cI$-Markov equivalent to $\cG_2$ and $\cG_3$.\qed
\end{example}

\subsection{Graphical representations for latent confounding}\label{subsec:graphical-unobserved}

Thus far, we have discussed how a structural causal model defines a \textit{data-generating process} for a particular system and interventions on that system.
In the simplest case, called the \textit{causally sufficient} setting, one directly observes the generated data.
However, it is often the case that observations are subject to additional processing, in which case we call the setting \textit{causally insufficient}.
Two forms of causal insufficiency are commonly considered.
First, under \textit{latent confounding}, some of the endogenous variables are simply unobserved, and we call these variables \textit{latent confounders}.
Thus, instead of observing samples from the distribution $\bbP_X$, one observes samples from a \textit{marginal} distribution $\bbP_{X'}$ for $X' \subset X$.
For instance, suppose that in \rref{ex:mouse-scm}, the experimentalist does not record the variable $X_1$ indicating whether the mice were genetically modified.
Then, an observer looking at their data would see samples from the distribution $\bbP_{X_2, X_3, X_4, X_5}$.
Second, under \textit{selection bias}, the probability that a sample is observed may depend on the values of some of the variables in the sample.
Thus, if we introduce a binary variable $S$ to indicate whether a sample is observed, and we have $\bbP(S = 1 \mid X)$ describe the selection process, then one observes samples from the \textit{conditional} distribution $\bbP(X \mid S = 1)$.
For instance, suppose that in \rref{ex:mouse-scm}, the experimentalist only records those experiments for which the mouse in the final generation weighs more than 20 grams.
Then, someone looking at their data would see samples from the distribution $\bbP_X(\cdot \mid X_5 \geq 20)$.

In this section, we will focus on the first type of causal insufficiency, latent confounding.
We postpone discussion of selection bias to \rref{sec:discussion}.
Without causal sufficiency, one must somehow account for latent confounders to perform accurate causal structure learning.
When the latent confounders have special structure, it may be possible to explicitly \textit{recover} the relationship of the latent confounders and the observed variables.
One such case is when each latent confounder is a parent of a large portion of the observed variables, which is termed \textit{pervasive confounding}.
In such settings, the observed data may be ``deconfounded" by removing its top principal components \cite{frot2017robust,shah2020right}, even when the causal relations are non-linear~\cite{agrawal2022Decamfounder}.
A large range of assumptions on the structure between the unobserved and observed variables may be suitable for different applications.
A thorough summary of methods using such assumptions is outside of the scope of the current review.
Instead, we focus on a different approach for accounting for latent confounders, which acknowledges their presence but does \textit{not} attempt to explicitly recover their relationships with the observed variables.

Structural assumptions on latent confounders can leave a wide range of signatures on the distribution of the observed variables.
These signatures include not only conditional independence constraints, which can be expressed in the form $\bbP_X(X_i, X_j \mid X_C) = \bbP_X(X_i \mid X_C) \bbP_X(X_j \mid X_C)$, but also more complex constraints.
This includes both equality constraints on the distribution $\bbP_X$, commonly called \textit{Verma constraints}, as well as inequality constraints.
The full set of constraints is referred to as a \textit{marginal DAG model} \cite{evans2016graphs}, and can be graphically modeled using a hypergraph.
Indeed, \cite{evans2016graphs} show that ordinary mixed graphs are incapable of representing marginal DAG models.
Nevertheless, ordinary mixed graphs are capable of encoding a rich subset of the constraints implied by a marginal DAG model.
For example, an \textit{acyclic directed mixed graph (ADMG)} encodes a subset of the equality constraints of the marginal DAG model via the associated \textit{nested Markov model} \cite{shpitser2014introduction,richardson2017nested}; in fact, the nested Markov model is known to encode \textit{all} equality constraints in the case of discrete variables \cite{evans2018margins}.
It is outside the scope of this review to provide a full overview of the different types of graphs used to capture the constraints of marginal DAG models, instead see \cite{evans2016graphs} and \cite{lauritzen2018unifying} for more thorough overviews.

In our review, we focus on (directed) \textit{ancestral graphs}, which encode only conditional independencies, are closed under marginalization, and have at most one edge between each pair of vertices.
Directed ancestral graphs are \textit{mixed} graphs, consisting of both directed and \textit{bidirected} edges.
A bidirected edge between two nodes indicates the possibility that they are both children of the same unobserved variable(s).
Similarly to directed graphs in the causally sufficient setting, the mixed graphs in the causally insufficient case are required to obey a form of acyclicity condition.
In particular, a mixed graph with directed and bidirected edges is called ``ancestral" if there are no directed cycles, and if any two nodes that are connected by a bidirected edge (called \textit{spouses}) are not ancestors of one another \cite{richardson2002ancestral}.

Similarly to DAG models, there is a notion of separation in directed ancestral graphs called \textit{m-separation}.
The same definition works as for d-separation: two nodes are \textit{m-connected} by a path $\gamma$ given a set of nodes $C$ if (1) every non-collider on the path is not in $C$, and (2) every collider on the path is either in $C$ or has a descendant in $C$.
Unfortunately, this notion of separation has the property that two non-adjacent nodes may have \textit{no} m-separating set.
Fortunately, adding a bidirected edge between two such nodes does not affect the set of m-separation statements which hold in the directed ancestral graph (\cite{richardson2002ancestral}, Theorem 5.1).
The operation of adding bidirected edges between all such nodes is called taking the \textit{maximal completion} of a directed ancestral graph, and a directed ancestral graph is called \textit{maximal} if it is its own maximal completion.
It is natural in structure learning to restrict the search space to directed maximal ancestral graphs (\textit{DMAGs}), so that each adjacency between nodes corresponds exactly to the lack of an m-separating set.

\begin{figure}
    \centering
    \includegraphics[width=.8\textwidth]{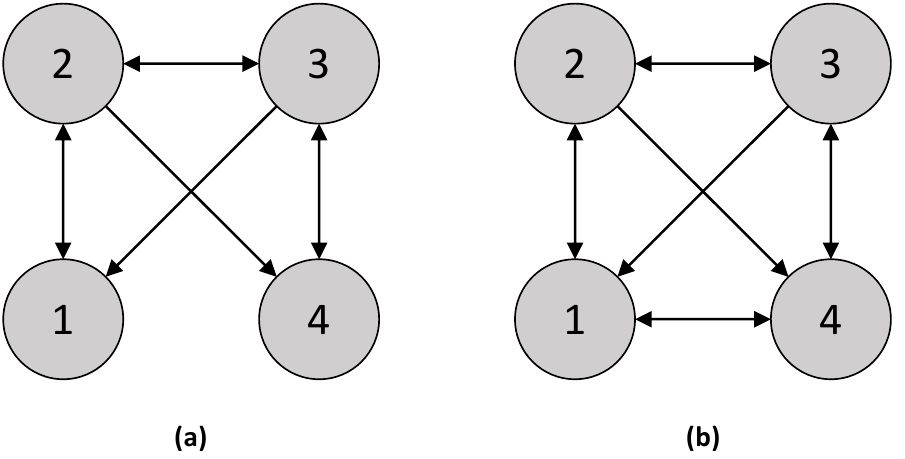}
    \caption{\textbf{(a)} An ancestral graph that is not maximal; \textbf{(b)} shows its maximal completion.}
    \label{fig:maximal-ancetral-graph}
\end{figure}

\begin{example}[Maximal completion]
\rref{fig:maximal-ancetral-graph} shows a graph (left) which is not maximal, since 1 and 4 are m-connected given any of the sets $\{ \varnothing, \{ 2 \}, \{ 3 \}, \{ 2, 3 \} \}$, but they are not adjacent.
The graph on the right is its maximal completion. \qed
\end{example}

\section{Identifiability}\label{sec:identifiability}

As alluded to in the previous section, two Markov equivalent DAGs cannot be distinguished from observational data alone.
In particular, given a DAG $\cG$, consider the collection of distributions $\bbM(\cG)$ which factorize according to $\cG$, i.e., can be written in the form \rref{eq:factorization}.
This collection depends on the allowed set of conditional distributions $\bbP_X(X_i \mid X_{\pa(i)})$.
If the set of conditional distributions is unrestricted, then we have that $\bbM(\cG) = \bbM(\cG')$ if and only if $\indepmodel(\cG) = \indepmodel(\cG')$, i.e., Markov equivalent DAGs give rise to the exact same set of distributions.
If the conditional distributions are restricted to specific classes, such as Gaussians or discrete measures, then this equivalence remains \cite{studeny2006probabilistic,maathuis2018handbook}.

Broadly speaking, there are two approaches to distinguishing between Markov equivalent DAGs.
The first approach, which we call the \textit{functional form} approach, considers restricting the class of conditional distributions in such a way that identifiability is possible from only observational data.
The second approach, which we call the \textit{equivalence class approach}, does not restrict the class of conditional distributions, but instead uses interventional data to refine the level of identifiability from the MEC to the $\cI$-MEC.
Given enough interventions, the equivalence class approach is sufficient for completely identifying a DAG or an ADMG \cite{eberhardt2005number}.

\subsection{Functional form approaches to identifiability}
Suppose the true causal graph is $X_1 \to X_2$. 
The core idea in this class of approaches is to find asymmetries between models learned in the ``causal" ($X_1 \to X_2$) and ``anticausal" ($X_2 \to X_1$) directions.
The asymmetries in this bivariate case are often easy to subsequently extend to the multivariate case.

As a canonical example, assume that noise is \textit{additive}, i.e., $X_2 = f_2(X_1) + \epsilon_2$, with $\epsilon_2 \indep X_1$.
By making assumptions about the functional form of $f_2$ and the distribution of $\epsilon_2$, it is often possible to show that the induced distribution $\bbP_X$ cannot be induced by a model of the form $X_1 = f_1(X_2) + \epsilon_1$, $\epsilon_1 \indep X_2$, under the same assumptions on $f_1$ and $\epsilon_1$.
For example, \cite{kano2003causal,shimizu2006linear,shimizu2011directlingam} assume that each function $f_i$ is linear, and each $\epsilon_i$ is non-Gaussian.
Indeed, \cite{hoyer2008nonlinear} shows that in linear models, symmetry is only possible in the Gaussian case, and give more general results for the case where $f_i$ is non-linear, which form the basis for structure learning methods such as the \textit{Causal Additive Model} (CAM) algorithm \cite{buhlmann2014cam}.
Even in the linear Gaussian case, it is possible to achieve identifiability by imposing additional assumptions, such as equal error variances for each $\epsilon_i$ \cite{peters2014identifiability}.
It is also possible to move beyond the additive noise case, e.g. by allowing for further nonlinearities after the addition of noise~\cite{zhang2009identifiability}.

Thus far, we have discussed identification strategies designed for continuous random variables.
Similar results are achievable in the discrete case, e.g. by assuming that the exogenous noise terms have low entropy \cite{kocaoglu2017entropic}, or by assuming the existence of a (hidden) low cardinality representation of the cause variable that mediates its effects \cite{cai2018causal}.

\subsection{The equivalence class approach to identifiability}

When no assumptions are made on the functional form, and only observational data is available, the true graph $\cG^*$ can only be identified up to the MEC, i.e., the set of DAGs $\cG'$ such that $\cG' \approx_\cM \cG^*$.
Thus, for the purposes of algorithm design, it becomes interesting to characterize when two DAGs are Markov equivalent.

\subsubsection{Characterizations of Markov equivalence classes}

\begin{figure}
    \centering
    \includegraphics[width=\textwidth]{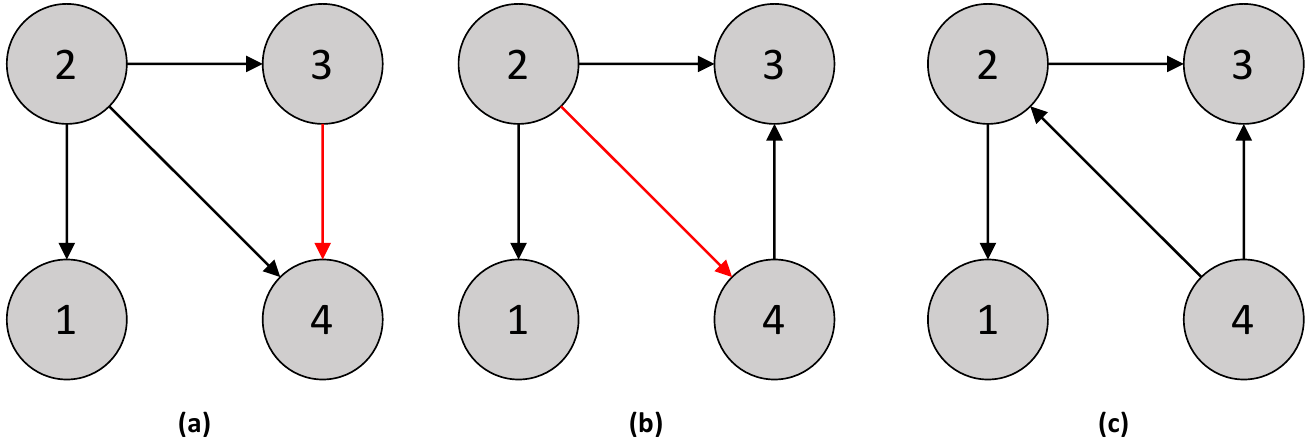}
    \caption{\textbf{Transformational characterization of equivalence in a DAG.} Starting with the DAG on the left, we proceed to the right by performing covered edge reversals on the red edges.}
    \label{fig:transformational-dag}
\end{figure}
\textbf{Characterizations of Markov equivalence in DAGs.}
There are numerous ways to characterize Markov equivalence in DAGs, and we will cover three main characterizations: a \textit{graphical} characterization, a \textit{transformational} characterization, and a \textit{geometric} characterization.
The graphical characterization is based on two notions.
The \textit{skeleton} of a DAG $\cG$ is defined as the set $\skel(\cG) = \{ (i, j) \mid i \to j ~\textrm{or}~ j \to i ~\textrm{in}~\cG \}$.
The v-structures (also called \textit{immoralities}) are defined as $\vstruct(\cG) = \{ (i, j, k) \mid i \to j \leftarrow k ~\textrm{in}~\cG, (i, k) \not\in \skel(\cG) \}$.
\cite{verma1990equivalence} shows that two DAGs $\cG$ and $\cG'$ are Markov equivalent if and only if they have the same skeleton and v-structures, i.e., $\cG \approx_\cM \cG'$ if and only if $\skel(\cG) = \skel(\cG')$ and $\vstruct(\cG) = \vstruct(\cG')$.
Given this graphical notion, it is natural to represent an MEC via an \textit{essential graph}, which is a mixed graph with the same adjacencies as all DAGs in the equivalence class, and with the edge $i \to j$ directed only if $i \to j$ in all DAGs in the equivalence class.
Meanwhile, the transformational characterization is based on a single notion: a \textit{covered edge} is an edge $i \to j$ in $\cG$ such that $\pa_\cG(i) = \pa_\cG(j) \setminus \{ i \}$.
$\cG'$ and $\cG$ are related by a \textit{covered edge flip} if $\cG'$ has all of the same edges as $\cG$, except that the covered edge $i \to j$ in $\cG$ is oriented as $j \to i$ in $\cG'$.
From the graphical characterization, one can deduce that if $\cG$ and $\cG'$ are related by a series of covered edge flips, then $\cG \approx_\cM \cG'$.
The transformational characterization states that the converse is also true: if $\cG \approx_\cM \cG'$, then $\cG$ can be transformed into $\cG'$ by a series of covered edge flips \cite{chickering1995transformational}.
This transformation is illustrated in \rref{fig:transformational-dag}.
Finally, the geometric characterization encodes each graph as an integer-valued vector in the space $\bbZ^{2^{[p]}}$.
First, we introduce a set of basis vectors $\delta_A$ for all subsets $A \subset [p]$.
Then, the \textit{standard imset} for a DAG $\cG$ is given by $u_\cG = \delta_{[p]} - \delta_{\varnothing} + \sum_{i=1}^p \left( \delta_{\pa_\cG(i)} - \delta_{\{ i \} \cup \pa_\cG(i)} \right)$.
Alternatively, \cite{studeny2010characteristic} introduces the \textit{characteristic imset} $c_\cG$, with $c_\cG(A) = 1$ if and only if there exists some $i \in [p]$ such that $A \setminus \{ i \} \subseteq \pa_\cG(i)$.
Two DAGs $\cG$ and $\cG'$ are Markov equivalent if and only if $u_\cG = u_{\cG'}$, or equivalently, $c_\cG = c_{\cG'}$.
\parspace

\textbf{Characterizations of interventional Markov equivalence in DAGs.}
As discussed in \rref{subsec:interventions}, the effect of an intervention can be formalized by introducing new binary variables to represent each intervention \cite{mooij2020joint}.
Therefore, the same characterizations of Markov equivalence that apply in the observational case just discussed also apply in the interventional case.
However, it is still instructive to directly characterize the interventional Markov equivalence class.
Consider a set of interventions $\cI$ such that $\varnothing \in \cI$ (i.e., observational data is available).
Extending a result for perfect interventions \cite{hauser2012characterization}, \cite{yang2018characterizing} shows that two DAGs $\cG$ and $\cG'$ are $\cI$-Markov equivalent if and only if they (1) have the same skeleton and v-structures, as in the case of a DAG and (2) for all $I \in \cI$ and $i \in I$, $j \not\in I$, we have $j \to i$ in $\cG$ if and only if $j \to i$ in $\cG'$.
Note that this is equivalent to stating that the two $\cI$-DAGs do not differ in v-structures of the form $\zeta_I \to i \leftarrow j$, confirming the equivalence with the observational characterization applied to $\cI$-DAGs.
As an example, under the set of interventions $\cI = \{ \varnothing, \{ 1 \} \}$, the graphs $\cG_2$ and $\cG_3$ in \rref{fig:markov-equivalence} are $\cI$-Markov equivalent, but $\cG_1$ is not, since its $\cI$-DAG would not have the v-structure $\zeta_{\{1\}} \to 1 \leftarrow 2$.
\parspace

\begin{figure}
    \centering
    \includegraphics[width=\textwidth]{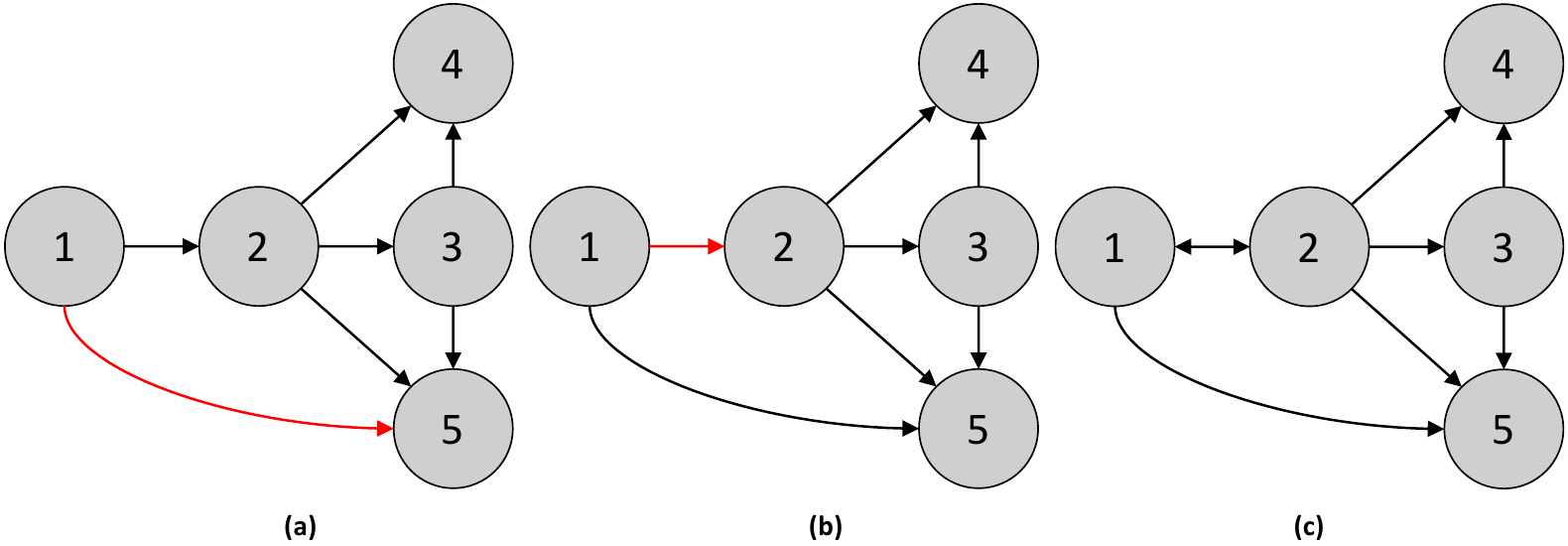}
    \caption{\textbf{Transformational characterization of equivalence in a DMAG.} Starting with the DMAG on the left, we proceed to the right by performing legitimate mark changes on the red edges.}
    \label{fig:transformational-mag}
\end{figure}

\textbf{Characterizations of Markov equivalence in DMAGs.}
As in the case of DAGs, equivalence between DMAG models can be characterized in multiple ways, and we will cover the graphical and transformational characterizations.
For both characterizations, we must define the notion of a \textit{discriminating path} for a vertex $k$.
A path $\gamma = \langle i, \ldots, k, j \rangle$ is a discriminating path for $k$ if (i) there is at least one node on the path between $i$ and $k$, (ii) every node between $i$ and $k$ is a collider on the path, and (iii) every node between $i$ and $k$ is a parent of $j$.
We denote the set of discriminating paths for node $k$ in $\cG$ as $\discr_k(\cG)$.
A fundamental result \cite{spirtes1996polynomial} states that two DMAGs $\cG$ and $\cG'$ are Markov equivalent if and only if (i) they have the same skeleton and v-structures, and (ii) for all $k$, for all $\gamma \in \discr_k(\cG) \cap \discr_k(\cG')$, $k$ is a collider on $\gamma$ in $\cG$ if and only if $k$ is a collider on $\gamma$ in $\cG'$.
Checking this graphical condition for Markov equivalence can be computationally expensive, motivating recent work \cite{hu2020faster} which provides a new graphical characterization of Markov equivalence in DMAGs that can be checked more efficiently.
We next describe the transformational characterization of Markov equivalence in DMAGs.
As in the case of DAGs, the transformational characterization requires us to define a local structural modification.
In particular, the modification of the edge $i \to j$ in $\cG$ to the edge $i \leftrightarrow j$ in $\cG'$, or vice versa, is called a \textit{legitimate mark change} \cite{zhang2005transformational} if (i) $\pa_\cG(i) \subseteq \pa_\cG(j)$, (ii) $\spo_\cG(i) \setminus \{ j \} \subseteq \spo_\cG(j) \cup \pa_\cG(j)$, and (iii) there is no $\gamma \in \discr_i(\cG)$ for which $j$ is the endpoint adjacent to $i$.
The authors in \cite{zhang2005transformational} show that $\cG \approx_\cM \cG'$ if and only if $\cG$ and $\cG'$ are connected by a series of legitimate mark changes.
This transformation is illustrated in \rref{fig:transformational-mag}.

\begin{table}[!b]
    \centering
    \begin{tabular}{|c|cccccccc|}
        \hline
        $p$ & 3 & 4 & 5 & 6 & 7 & 8 & 9 & 10 
        \\
        \hline
        \# MEC & 11 & 185 & 8.78e4 & 1.06e6 & 3.13e8 & 2.12e11 & 3.26e14 & 1.12e18
        \\
        $\frac{\textrm{\# MEC}}{\textrm{\# DAG}}$ & 0.44 & 0.34 & 0.30 & 0.28 & 0.27 & 0.27 & 0.27 & 0.27
        \\
        $\frac{\textrm{\# MEC-1}}{\textrm{\# MEC}}$ & 0.36 & 0.32 & 0.30 & 0.29 & 0.28 & 0.28 & 0.28 & 0.28
        \\
        \hline
    \end{tabular}
    \caption{The number of MECs (first row), the ratio of the number of MECs to the number of DAGs (second row), and the ratio of the number of MECs of size 1 compared to the total number of MECs (third row), up to 10 nodes.}
    \label{tab:number-mecs}
\end{table}

\subsubsection{Combinatorial aspects of Markov equivalence}

Since DAGs in general can only be identified up to ($\cI$-)Markov equivalence, it has been of significant interest to study the size of a given MEC, the number of MECs over a given number of variables, and the minimum number of interventions required to identify a DAG (i.e., obtain a $\cI$-MEC of size 1).

The first problem - computing the number of DAGs within a given MEC, or computationally equivalently, sampling uniformly from the MEC - is important for a number of experimental design algorithms \cite{ghassami2018budgeted}, which use Monte-Carlo approximations to compute expectations over the MEC and pick interventions with good average-case behavior.
A recent advance \cite{wienobst2021polynomial} provides a polynomial-time algorithm for this task based on a representation of the equivalence class via clique trees, improving over previous algorithms with exponential worst-case runtime \cite{he2015counting,bernstein2017sampling,talvitie2019counting,ghassami2019counting,ahmaditeshnizi2020lazyiter,ganian2022efficient}.

To address the second problem, \cite{gillispie2001enumerating} develops a program for enumerating all MECs on graphs with a given number of nodes, and obtained results for graphs of up to 10 nodes, shown in \rref{tab:number-mecs}.
Further theoretical works \cite{radhakrishnan2017counting,radhakrishnan2018counting} study the problem of enumerating all MECs for a fixed skeleton using the idea of generating functions from combinatorics.
The computational results in \cite{gillispie2001enumerating} suggest that, asymptotically, the average MEC contains approximately 4 DAGs, and that roughly one quarter of all MECs are comprised of only a single DAG, in which case no interventional data is needed to identify the causal DAG.
However, proving these conjectured limits, as well as efficiently enumerating the number of MECs on a given number of nodes, remain open combinatorial and computational problems.
Less work has been done to characterize the average number of interventions required to identify a DAG.
For a given DAG, \cite{squires2020active} characterizes the minimum-size set of single-node interventions needed to identify the underlying causal DAG, using a representation based on clique trees.
However, this work does not address the average of this quantity over all DAGs on a given number of nodes.
Meanwhile, \cite{katz2019size} conducts a computational study of the average number of \textit{greedily} selected interventions to identify a graph, where the average is with respect to a directed Erd{\"o}s-R{\'e}nyi graph model.
In this model, the results suggest that the number of interventions necessary is typically less than 4, but further work is necessary to characterize the average with respect to the uniform distribution over graphs and to address the case where interventions are picked optimally.

\section{Methods for Causal Structure Learning}\label{sec:algorithms}

Thus far, we have discussed what is in principle identifiable about the underlying causal DAG with observational and interventional data.
Now, we present algorithms which carry these principles of identifiability into practice.
In particular, we will discuss a number of algorithms which are \textit{consistent}, i.e., in the limit of infinite data, they provably learn all identifiable causal structures.
We will also highlight some \textit{heuristic} algorithms, which do not have consistency guarantees but often perform well in practice.
We begin with a broad overview of the different paradigms for causal structure learning, before diving into methods which explicitly leverage the combinatorial structures already discussed.
At the highest level, methods for estimating causal models from data fall into two broad categories: \textit{constraint-based} methods and \textit{score-based} methods.
Constraint-based methods are natural when viewing causal structure learning as a constraint satisfaction problem, where conditional independences or other constraints that can be inferred from data are used to iteratively prune the space of possible graphs.
In contrast, score-based methods arise from viewing causal structure learning as a \textit{combinatorial optimization} problem. 
These methods assign a score to each graph (or equivalence class) which quantifies how well it fits the data, then search the space of graphs (or equivalence classes) to find a model which optimizes the score.
To highlight the general principles of these two paradigms, we will first concentrate on the causally sufficient case with only observational data.
Then, in \rref{subsec:causal-structure-learning-interventional}, we discuss algorithms that can make use of interventional data, and in \rref{subsec:causal-structure-learning-confounding}, we briefly discuss algorithms for learning in the presence of latent confounding.
\parspace

\textbf{Constraint-based approaches.} The most prominent constraint-based approach to causal structure learning is the PC algorithm \cite{spirtes2000causation,kalisch2007estimating}.
The PC algorithm begins with a complete undirected graph, and iteratively deletes edges by testing conditional independences involving conditioning sets of increasing cardinality.
Then, the second phase of the PC algorithm orients v-structures by re-using the conditional independences found in the first phase.
Additional orientations can be inferred via the \textit{Meek orientation rules} \cite{meek1995causal}.

The method for testing conditional independence (CI) depends on modeling assumptions as well as practical considerations such as computational complexity.
For example, in a multivariate Gaussian distribution, two variables $X_i$ and $X_j$ are conditionally independent given the variables $X_C$ if and only if the partial correlation $\rho_{ij \mid C}$ is zero.
Since the distribution of sample partial correlation coefficients is well-known (see e.g. \cite{kalisch2007estimating}), hypothesis testing for CI in the Gaussian setting is straightforward and computationally efficient.
On the other hand, in non-parametric settings, hypothesis tests for conditional independence can often be performed based on more complicated test statistics \cite{zhang2011kernel,heinze2018invariant,strobl2019approximate}.
Unfortunately, impossibility results \cite{shah2020hardness} state that any \textit{uniformly} valid conditional independence test (i.e., one whose false positive rate tends to at most the significance level $\alpha$, over all possible distributions $\bbP$ where $X \indep_\bbP Y \mid Z$) will have no statistical power (i.e., the probability of a true positive will also be at most $\alpha$).
Thus, testing conditional independence requires additional assumptions on the set of possible distributions, such as complexity restrictions on the function space of $\bbE_\bbP [X \mid Z]$.

Under such complexity assumptions, conditional independence tests allow constraint-based approaches to directly be applied to non-parametric settings, even permitting high-dimensional consistency bounds in these settings \cite{harris2013pc}.
Furthermore, because conditional independences also characterize DMAG models, constraint-based approaches can be easily extended to settings with latent variables \cite{colombo2012learning}.
Pushing further, one may encode conditional independences as logical constraints, allowing them to be used in answer set programming (ASP) solvers.
These solvers can search over more general model classes and easily incorporate background knowledge \cite{hyttinen2014constraint,zhalama2017sat}.
However, ASP-based causal structure learning methods are widely viewed as being difficult to scale for many practical applications.
\parspace

\textbf{Score-based approaches.} Score-based methods for causal structure learning originated in parametric settings, such as in discrete or linear Gaussian models.
In parametric settings, the score $S(\cG)$ of a graph $\cG$ is often based on the \textit{marginal likelihood} $\bbP(\bbX \mid \cG)$ of the data $\bbX$ given the graph $\cG$, with respect to some prior $\bbP(\theta)$ over the parameters $\theta$.
In some cases, e.g. when choosing a \textit{conjugate} prior for the likelihood function, $\bbP(\bbX \mid \cG)$ can be computed in closed form \cite{geiger2002parameter}.
Alternatively, it is common to use a consistent approximation of the marginal likelihood, in the form of the \textit{Bayesian information criterion (BIC)} score \cite{chickering2002optimal,chickering2020statistically}.
Such likelihood-based scores can be extended to nonparametric settings, e.g. by using Gaussian process priors \cite{friedman2000gaussian} or nonparanormal distributions \cite{nandy2018high}.
The BIC score and related scores are also a natural starting point from which to develop more sophisticated scores with better statistical and computational properties, see e.g. \cite{brenner2013sparsityboost}.

Finding the highest-scoring DAG model is generally NP-hard \cite{chickering1996learning}, imposing a tradeoff between computational efficiency and algorithmic consistency guarantees.
Score-based methods can generally be subdivided into three categories based on how they address this tradeoff.
On one end of the spectrum, \textit{exact} score-based approaches find some $\widehat{\cG}$ that exactly optimizes the score $S$.
Exact approaches address computational issues using a variety of combinatorial optimization techniques and heuristics, e.g., dynamic programming \cite{koivisto2004exact,parviainen2010bayesian}, A*-style state-space search \cite{yuan2013learning}, or methods from integer linear programming \cite{jaakkola2010learning,cussens2011bayesian,bartlett2017integer,cussens2020gobnilp}.
For example, the GOBNILP algorithm \cite{cussens2020gobnilp} uses the geometric characterization of Markov equivalence classes to reduce structure learning to an integer linear programming problem.
This reduction allows the use of techniques such as cutting planes and pricing to handle the exponential number of decision variables and constraints.

\textit{Greedy} score-based approaches trade off to achieve better computational efficiency over exact approaches by relaxing the requirement that $\widehat{\cG}$ optimizes $S$.
Most prominently, Greedy Equivalence Search (GES) \cite{chickering2002optimal} and its variants \cite{chickering2020statistically} perform a search over equivalence classes of graphs that greedily optimizes $S$.
While greedy algorithms are not \textit{exact}, they are still \textit{consistent}, placing them in a middle ground on the computational-statistical tradeoff.
Notably, \cite{linusson2021greedy} shows that GES and a number of other greedy approaches can also be viewed geometrically.
In particular, these methods can be seen as \textit{edge walks} between vertices of the \textit{characteristic imset polytope}, i.e., the convex hull of all characteristic imsets on $p$ variables.
Finally, at the other extreme of this tradeoff, gradient-based methods \cite{zheng2018dags,yu2019dag,lachapelle2019gradient,zhang2019d} relax the discrete search space over DAGs to a \textit{continuous} search space, allowing gradient descent and other techniques from continuous optimization to be applied to causal structure learning.
However, the search space of these problems is highly non-convex, so that the the optimization procedure may become stuck in a local minima.
Thus, consistency guarantees for these methods will depend on theoretical advances in global minimization of such non-convex optimization problems.

\subsection{Learning DAGs using permutation-based algorithms}

Beyond the constraint-based and score-based paradigms for causal structure learning already discussed, there are a variety of \textit{hybrid} methods \cite{tsamardinos2006max,schmidt2007learning,schulte2010imap,alonso2013scaling,nandy2018high}, which generally use constraints to reduce the search space, and scores to optimize over this reduced search space.
In this section, we discuss the \textit{greedy sparsest permutation} (GSP) algorithm, a hybrid method that constrains the search space to the set of (estimated) minimal I-MAPs of $\bbP_X$.
By focusing on this method, we highlight the combinatorial nature of the problem of causal structure learning.

As discussed in \rref{sec:preliminaries}, a distribution $\bbP_X$ may permit several different minimal I-MAPs.
Since the minimal I-MAPs of $\bbP_X$ are the (locally) sparsest DAGs which can correctly model $\bbP_X$, they form a natural space over which to search for the true DAG $\cG^*$.
Furthermore, the space of minimal I-MAPs of $\bbP_X$ can be described as the image of a $\bbP_X$-dependent map, with the $\bbP_X$-independent domain of $S_p$ of \textit{permutations} of $[p]$.
We denote by $i <_\pi j$ that $i$ is earlier in the permutation $\pi$ than $j$, and we call a graph $\cG$ \textit{consistent} with a permutation $\pi$ if and only if $i <_\pi j$ implies that $j \not\to i$ in $\cG$.
The following result establishes the existence of a unique map from permutations to minimal I-MAPs.
\begin{thm}[from \cite{verma1990causal}]\label{thm:minimal-imap}
Given a permutation $\pi$ and a distribution $\bbP_X$, there exists a unique graph $\cG_{\bbP_X}(\pi)$ that is consistent with $\pi$ and is a minimal I-MAP for $\bbP_X$.
This graph has edges
\[
\{ i \to j \mid X_i \not\indep X_j \mid X_{\pre_\pi(j) \setminus \{ i \}} \}
\quad\quad
\textrm{where}~
\pre_\pi(j) = \{ k \mid k <_\pi j \}.
\]
\end{thm}

\begin{figure}
    \centering
    \includegraphics[width=.99\textwidth]{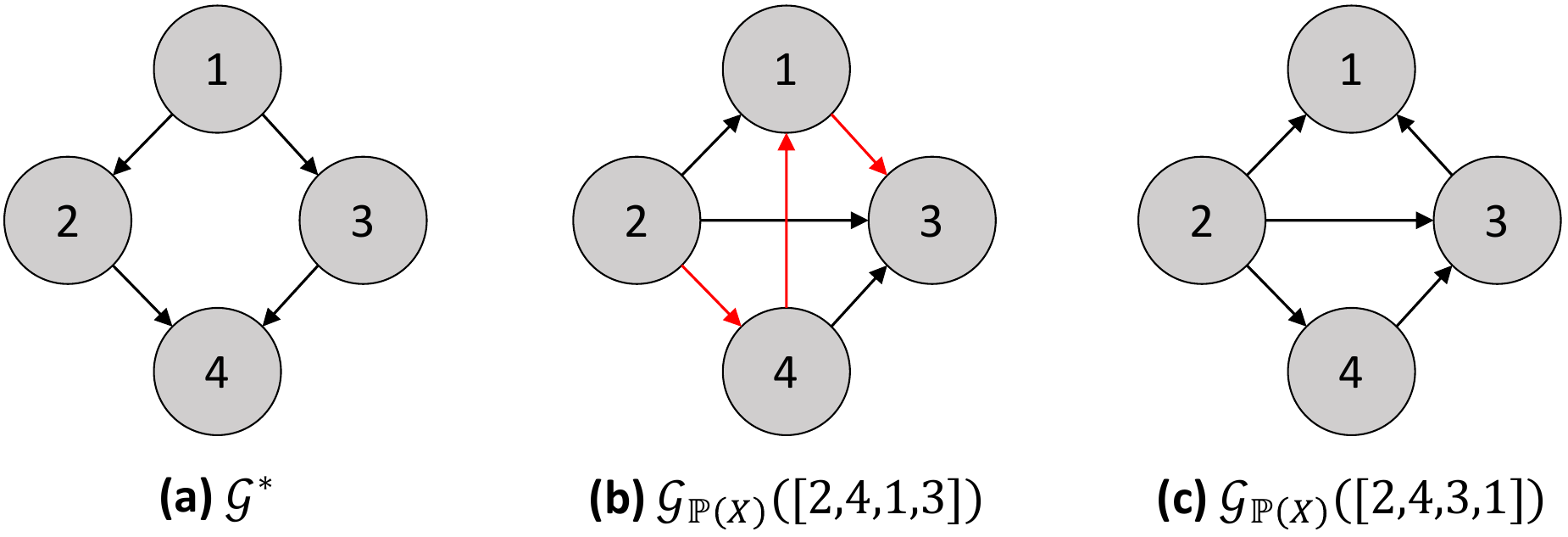}
    \caption{
    \textbf{A greedy step over minimal I-MAPs performed by GSP.} 
    (a) The true graph $\cG^*$, to which the distribution $\bbP(X)$ is faithful.
    (b) The minimal I-MAP associated with the permutation $\pi^{(0)} = [2,4,1,3]$, with covered edges shown in red.
    (c) The minimal I-MAP associated with the permutation $\pi^{(1)} = [2,4,3,1]$, obtained after flipping the covered edge $1 \to 3$.
    }
    \label{fig:gsp-minimal-imap}
\end{figure}

Given a graph $\cG$, let $|\cG|$ be the number of edges in the graph.
The \textit{sparsest I-MAP theorem} \cite{raskutti2018learning} establishes that, under a mild condition, the sparsest minimal I-MAPs of $\bbP_X$ - i.e, those such that $|\cG_{\bbP_X}(\pi)|$ is minimized - are Markov equivalent to the underlying causal graph $\cG^*$.
In particular, the required condition for this result is strictly weaker than the \textit{restricted faithfulness assumption} \cite{ramsey2006adjacency}, which only requires that $\indepmodel(\bbP_X)$ and $\indepmodel(\cG^*)$ agree on conditional independences/d-separations involving nodes connected by paths of lengths one or two.
The sparsest I-MAP theorem directly suggests the \textit{sparsest permutation} (SP) algorithm: enumerate over all permutations $\pi \in S_p$, estimating the minimal I-MAP $\cG_{\bbP_X}(\pi)$ for each of these permutations using conditional independence testing, and return the sparsest graphs.

However, the SP algorithm is clearly computationally prohibitive, since the size of $S_p$ is super-exponential in $p$.
To address this issue, \cite{solus2020consistency} proposed the \textit{greedy sparsest permutation} (GSP) algorithm.
GSP searches greedily over the space of permutations, and hence, minimal I-MAPs.
In particular, at each step $i$ of the algorithm, GSP maintains a permutation $\pi^{(i)}$ and its corresponding minimal I-MAP $\cG_{\bbP_X}(\pi^{(i)})$.
At this step, GSP searches over the Markov equivalence class of $\cG_{\bbP_X}(\pi^{(i)})$ for some DAG $\cG'$ which is \textit{not} a minimal I-MAP of $\bbP_X$.
This search can be executed by repeatedly flipping covered edges to generate new permutations.
Upon finding $\cG'$ which is not a minimal I-MAP of $\bbP_X$, there must be some strict sub-DAG $\cG''$ of $\cG'$ which is a minimal I-MAP of $\bbP_X$.
GSP then takes the topological ordering of this sub-DAG as the new permutation $\pi^{(i + 1)}$, with $\cG''$ as its corresponding minimal I-MAP $\cG_{\bbP_X}(\pi^{(i + 1)})$.
One greedy step of GSP is demonstrated in \rref{fig:gsp-minimal-imap}.

As in the case for other greedy approaches, GSP has an interpretation as an edge walk over a convex polytope.
In particular, starting from the \textit{permutahedron}, i.e., the convex hull of all permutations on $p$ nodes, we may define the \textit{DAG associahedron} by contracting all edges $\pi^{(i)} - \pi^{(j)}$ of the permutahedron for which $\cG_{\bbP_X}(\pi^{(i)}) = \cG_{\bbP_X}(\pi^{(j)})$.
As shown in \cite{solus2020consistency}, this contraction results in a convex polytope, GSP is equivalent to an edge walk along this polytope, and, under conditions that are strictly weaker than the faithfulness assumption, this edge walk terminates in the Markov equivalence class of the causal graph $\cG^*$ underlying $\bbP_X$.
The central technical ingredient in this proof is the existence of \textit{Chickering sequences}.
In particular, \cite{chickering2002optimal} proves the \textit{Meek conjecture} for DAGs \cite{meek1997graphical}: if $\cG_M$ is an I-MAP of $\cG_0 = \cG^*$, then there exists a sequence $(\cG_0, \cG_1, \ldots, \cG_M)$ composed only of edge additions and covered edge reversals.
This sequence is called a Chickering sequence \cite{solus2020consistency} and its existence guarantees the consistency of GSP.

In addition to the consistency of GSP and the algorithms discussed previously, which provides guarantees as the sample size goes to infinity, it is important to understand the performance of different algorithms for finite sample size.
Simulation results suggest that score-based and hybrid approaches perform better for fixed sample sizes \cite{nandy2018high,heinze2018causal,andrews2019learning}.
However, a theoretical characterization of the trade-offs between these algorithms on finite samples is not well understood, and is an important area for future research, as also briefly described in \rref{sec:discussion}.

\subsection{Bayesian methods for causal structure learning}

Thus far, we have only discussed causal structure learning methods which return a \textit{point estimate} - i.e., a single DAG that (approximately or locally) maximizes a score, and/or satisfies inferred conditional independences.
However, when the amount of data is small, there may be substantial \textit{uncertainty} about the underlying graph (or equivalence class).
A common framework for quantifying this uncertainty is \textit{Bayesian inference}.
Given some dataset $\bbD$, instead of returning a point estimate, Bayesian methods return (an approximation to) the posterior $\bbP(\cG \mid \bbD)$ over graphs.
This posterior allows one to compute marginal probabilities of any feature of interest, such as the posterior probability of some edge $i \to j$.

Bayesian methods for causal structure learning can be divided into three types of approaches: \textit{exact} approaches (e.g., \cite{eaton2007exact}) and two types of approximate approaches: \textit{variational} and \textit{sampling-based} approaches.
Similarly to the gradient-based approaches discussed before, variational approaches do not necessarily return a consistent estimate of the posterior; rather, they project the posterior onto a \textit{variational family} $\{ Q(\cdot \mid \theta) \}_{\theta \in \Theta}$, which is more computationally convenient.
However, traditional variational families, such as multivariate Gaussians, are continuous and thus do not apply to the discrete setting of DAGs.
Thus, until recently, variational methods for Bayesian causal structure learning have not been widely studied.
For a recent work in this space, see \cite{lorch2021dibs}, which uses relaxations of DAGs to a continuous search space and neural networks to parameterize a flexible variational family.

On the other hand, sampling-based approaches to Bayesian causal structure learning have been much more widely studied.
Markov-chain Monte Carlo (MCMC) methods have been especially popular, beginning with the \textit{structure MCMC} algorithm \cite{madigan1995bayesian}, which runs a Metropolis-Hastings algorithm over the space of DAG models, using edge additions and deletions to move in this space.
However, this approach suffers from slow \textit{mixing times} due to regions of high-probability DAG models being separated by large regions of low-probability DAG models, i.e., if structure MCMC finds some high-probability DAG $\cG_0$, some other high-probability DAG $\cG_M$ may only be reachable from $\cG_0$ by a sequence of DAGs $\cG_1, \ldots \cG_{M-1}$ which have very low probability.
Thus, the probability that structure MCMC traverses this path becomes incredibly low, so that $\cG_M$ will not be sampled without running the algorithm for many steps.

This difficulty has motivated a search for ``smoother" sampling spaces, either by adding moves to structure MCMC \cite{giudici2003improving,grzegorczyk2008improving}, or by changing the search space, as was done in \textit{order MCMC} \cite{friedman2003being,ellis2008learning}, \textit{partial order MCMC} \cite{niinimaki2016structure}, and \textit{partition MCMC} \cite{kuipers2017partition}.
These methods run a Markov chain over some ``coarser" space (permutations, partial orders, or ordered partitions), then sample DAGs conditionally based on their consistency with the coarser structure.
The \textit{minimal I-MAP MCMC} algorithm \cite{agrawal2018minimal} also runs a Markov chain over the coarser space of permutations.
However, instead of conditionally sampling a DAG based on each permutation, it estimates the minimal I-MAP associated to each sampled permutation.

Since the space of permutations is much smaller than the space of DAGs or MECs, the minimal I-MAP MCMC algorithm can mix more quickly than previous algorithms.
But this comes at a price: minimal I-MAP MCMC does not sample over the entire posterior distribution of DAG models, but only a restricted subset.
Luckily, this price is small:
intuitively, conditional on an order, the minimal I-MAP asymptotically has the highest posterior probability, so a point mass on the minimal I-MAP is a good approximation of the true conditional distribution.
Indeed, \cite{agrawal2018minimal} shows that the posterior approximation error for any bounded function of the graph decreases exponentially with the number of samples.
By highlighting this algorithm, we once again see the computational benefits that are possible when considering the combinatorial nature of the causal structure learning problem.

%

\subsection{Causal structure learning using interventional data}\label{subsec:causal-structure-learning-interventional}

As discussed in \rref{sec:identifiability}, interventional data can significantly improve the identifiability of causal models.
Several approaches have been proposed for learning from a combination of observational and experimental data, going back at least to the Bayesian approaches of \cite{cooper1999causal} and \cite{eaton2007exact}.
As in the case of learning from purely observational data, these approaches can be divided into constraint-based approaches, such as the COmbINE \cite{triantafillou2015constraint} algorithm, and score-based approaches.
Score-based approaches include greedy algorithms, such as \textit{Greedy Interventional Equivalence Search (GIES)} \cite{hauser2012characterization}, and gradient-based algorithms, such as meta-learning approaches \cite{ke2019learning} and DCDI \cite{brouillard2020differentiable}.
Note that, unlike in the case of GES for observational data, GIES is known to not be consistent for interventional data \cite{wang2017permutation}.

The Joint Causal Inference framework \cite{mooij2020joint} discussed in \rref{subsec:interventions} suggests a natural way to extend causal structure learning algorithms for observational data to settings with interventional data.
In particular, an algorithm for the observational setting can be used to learn the $\cI$-DAG by appending indicator variables to the dataset for each intervention $I \in \cI$, as long as the algorithm can incorporate appropriate forms of background knowledge.
This background knowledge includes \textit{exogeneity} - i.e., intervention variables are not caused by the original ``system" variables, \textit{randomized context} - i.e., lack of confounding between the intervention and system variables, and \textit{generic context} - i.e., that the intervention variables are deterministically related to one another.
As an example, \cite{squires2020permutation} shows that the GSP algorithm can be adapted to include these assumptions, along with any assumptions about known targets of each intervention, while maintaining consistency of the algorithm.
They call the resulting algorithm the \textit{Unknown Target Intervention GSP (UT-IGSP)} algorithm to emphasize its ability to handle interventions with unknown targets, extending previous works where targets were assumed to be known \cite{wang2017permutation,yang2018characterizing}.
Finally, it is also natural to develop Bayesian variants of causal structure learning algorithms for interventional data, e.g., \cite{castelletti2022network} shows how to compute posteriors over DAGs in the setting when the data is multivariate Gaussian.

\subsection{Causal structure learning in the presence of latent confounding}\label{subsec:causal-structure-learning-confounding}

The approaches to causal structure learning in the causally insufficient setting follow the same broad categorization as approaches in the causally sufficient setting.
In particular, the \textit{Fast Causal Inference (FCI)} algorithm \cite{spirtes1995causal} is a constraint-based algorithm for learning DMAGs, similar in spirit to the PC algorithm.
The FCI algorithm has inspired several variants, including \textit{Really Fast Causal Inference (RFCI)} \cite{colombo2012learning}, and \textit{FCI+} \cite{claassen2013learning}.
Score-based methods include both greedy search strategies, such as \textit{Greedy FCI (GFCI)} \cite{ogarrio2016hybrid}, \textit{MAG Max-Min Hill Climbing (M\textsuperscript{3}HC)} \cite{tsirlis2018scoring}, and \textit{Conservative rule and Causal effect Hill Climbing (CCHM)} \cite{chobtham2020bayesian}, exact score-based approaches, such as AGIP \cite{chen2021integer}, and gradient-based approaches \cite{bhattacharya2021differentiable}.

As in the case of learning DAGs, we will discuss a hybrid method for learning DMAGs, which combines elements of both score-based and constraint-based approaches, and elucidates the combinatorial aspects of learning DMAGs. This method, called the \textit{Greedy Sparsest Poset} (GSPo) method, restricts the search space of DMAGs to minimal I-MAPs of the distribution $\bbP_X$.
This space can be realized as the image of a map $\cG_{\bbP_X}$ from \textit{partially ordered sets (posets)} to graphs.
A partially ordered set $\pi$ defines a relation $\preceq_\pi$ that captures the notion of an ordering via three requirements: reflexivity ($i \preceq_\pi i$ for all $i$), antisymmetry ($i \preceq_\pi j$ and $j \preceq_\pi i$ implies $i = j$), and transitivity ($i \preceq_\pi j$ and $j \preceq k$ implies $i \preceq_\pi k$).
Because of the definition of the ancestrality condition, the set of complete DMAGs can be put in bijection to the set of posets, so that posets form a natural domain for the map $\cG_{\bbP_X}$.

The authors in \cite{bernstein2020ordering} show that $\cG_{\bbP_X}(\pi)$ can be constructed using a procedure similar to the procedure defined for DAGs in \rref{thm:minimal-imap}, although the construction requires two iterations of conditional independence testing between pairs of variables instead of one.
They also provided a version of the sparsest I-MAP theorem for DMAGs, i.e., under a restricted faithfulness assumption, the sparsest minimal I-MAPs of $\bbP_X$ are all Markov-equivalent to the underlying DMAG $\cG^*$.
Motivated by the GSP algorithm for learning DAGs, \cite{bernstein2020ordering} introduce the \textit{greedy sparsest poset (GSPo)} algorithm for learning DMAGs, which uses legitimate mark changes to search over posets and iteratively find sparser I-MAPs.
Over 100,000 synthetic examples suggest that the GSPo algorithm is consistent, but proof of its consistency is an important open problem, and closely tied to the open problem of generalizing Meek's conjecture \cite{meek1997graphical,chickering2002optimal} to DMAGs.

\section{Discussion and open problems}\label{sec:discussion}

In this review article, we sought to cover both classical and recent approaches to causal structure learning, emphasizing the combinatorial nature of this problem.
We end by discussing several related areas of work that were not covered in depth and remain under active development.

\textbf{\textit{Learning with both interventions and latent confounding.}}
While we separately discussed learning with interventional data and learning under confounding, it is natural to combine these two settings.
Recent work \cite{jaber2020causal} considers this combination for DMAGs, introducing the new notion of $\Psi$-Markov equivalence to capture pairs of graphs and interventions which induce the same set of conditional independencies and conditional invariances.
This work allows for both \textit{soft} and \textit{unknown-target} interventions.
Furthermore, \cite{jaber2020causal} provides a graphical characterization of $\Psi$-Markov equivalence, and introduces a constraint-based algorithm, called $\Psi$-FCI, for learning the $\Psi$-Markov equivalence class from data.
As a next step it is natural to consider score-based algorithms, both exact and greedy, for learning DMAGs, ADMGs, and other subclasses of marginal DAG models, using a combination of observational and interventional data.

\textbf{\textit{Learning with assumptions on the latent structure.}}
As indicated in \rref{subsec:graphical-unobserved}, in some cases with unobserved confounding, it is desirable to \textit{recover} the unobserved variables and their relationship to the observed variables.
Naturally, recovery of these details requires assumptions on their structure.
A common assumption, called the \textit{exogeneity} or \textit{measurement} assumption, is that all unobserved variables are upstream of the observed variables, i.e., none of the unobserved variables are caused by any of the observed variables.

With the exogeneity assumption as a starting point, additional assumptions may be made to (approximately) recover the latent variables, and possibly, the structure between them.
For example, several works \cite{shah2020right,frot2017robust} consider recovering the unobserved variables in settings with \textit{pervasive confounding}, i.e., when each unobserved variable has a direct effect on a large number of observed variables.
As an important special case of this setting, some works have considered recovering a \textit{mixture} of DAG models \cite{strobl2019global,strobl2019improved,saeed2020causal,gordon2021identifying}, where there is a single unobserved variable that is a parent of all variables in the graph.
Alternatively, many works \cite{jernite2013discovering,halpern2015anchored,kummerfeld2016causal,zhang2017causal_measurement,cai2019triad,xie2020generalized,kivva2021learning,saeed2020anchor} consider recovering unobserved variables under the measurement assumption and a form of \textit{purity} or \textit{anchor} assumption, where each unobserved variable must have some number of observed variables which are only their children.
Few works consider recovering unobserved variables \textit{without} the assumption of exogeneity, with \cite{squires2022causal} being a recent exception.

\textbf{\textit{Learning in the presence of selection bias.}}
As suggested in \rref{subsec:graphical-unobserved}, considerable effort has gone into characterizing the distributional constraints imposed by marginalization of DAG models.
However, in many applications, the observed distribution is the result of both marginalization and \textit{conditioning} of an underlying distribution.
In particular, such observed distributions are induced by \textit{selection bias}, where the probability that a sample is observed is dependent on the value of some of the variables in the sample.
General maximal ancestral graphs (see \rref{subsec:graphical-unobserved}), which allow for undirected edges in addition to directed and bidirected edges, are conditional independence models which are closed under marginalization and conditioning.
As in the case of marginalization, several graphical representations, including MC graphs \cite{koster2002marginalizing}, summary graphs \cite{wermuth2011probability}, and regression graphs \cite{wermuth2014graphical,sadeghi2016pairwise} have been introduced to capture constraints induced by such conditional models;
\cite{evans2016graphs} provides an exact graphical characterization of all equality and inequality constraints induced by marginalizing a DAG model, in the form of a hypergraph called an \textit{mDAG}.
In contrast, to the best of our knowledge, there is no graphical representation that exactly captures all constraints induced by conditioning a DAG model, though initial steps towards such a characterization are made in \cite{armen2018towards}.
In the special case of discrete random variables, \cite{lauritzen1999generating} shows that DAG models under selection are equivalent to hierarchical log-linear models, and \cite{evans2015recovering} provides graphical conditions under which the full DAG model is identifiable from the DAG model under selection.
To better understand DAG models under selection, next steps include (1) developing a graphical representation that fully captures both marginalization and conditioning, (2) developing notions of Markov equivalence in this setting, including with interventional data, and (3) developing structure learning algorithms in this general setting.
However, as noted by \cite{evans2016graphs}, the complexity of the inequality constraints introduced by general DAG models under marginalization and selection may render them too burdensome for practical use.
Therefore, it is of interest to characterize when simpler models such as MC graphs, regression graphs, and summary graphs are sufficient for downstream tasks.

\textbf{\textit{Learning cyclic causal models.}}
As indicated in \rref{sec:preliminaries}, a widespread assumption in causal modeling and causal structure learning is that the structural causal model (SCM) induces an acyclic graph.
However, this may not be the case if the SCM models a system that involves feedback loops.
While the underlying dynamics of the system are necessarily acyclic over time, feedback loops can arise when modeling the equilibrium states of such systems~\cite{bongers2018causal}.
For example, in gene regulatory networks, we may have that gene A regulates gene B, and gene B also regulates gene A, so that intervening on either gene will affect the value of the other gene.
Recent work \cite{bongers2021foundations} has investigated the semantics of cyclic causal models, showing that Markov properties and other desirable properties hold in the case of certain solvability conditions.
Despite the technical difficulties associated with cyclic models, several approaches have been proposed for learning their structure from data.
These approaches include many algorithms designed for the linear case, including LLC \cite{hyttinen2012learning}, score-based approaches \cite{ghassami2020characterizing}, and BackShift \cite{rothenhausler2015backshift}.
Algorithms for the general case  include SAT-based approaches \cite{hyttinen2013discovering},  exact score-based approaches \cite{rantanen2020discovering}, and constraint-based approaches \cite{forre2018constraint,mooij2020constraint,strobl2019constraint}.

\textbf{\textit{Statistical and computational complexity of causal structure learning.}}
In conjunction with methodological developments for settings with cycles, latent confounding, selection bias, and interventional data, it is important to understand the fundamental statistical and computational limits of causal structure learning, and any tradeoffs between these.
The analysis of existing causal structure learning algorithms gives \textit{upper bounds} on what is statistically and computationally achievable.
Recent work derives upper bounds for a wide range of settings, including the linear equal-variance setting \cite{ghoshal2017learning}, the linear non-Gaussian setting \cite{wang2020high}, other parametric settings \cite{park2019identifiability,rajendran2021structure}, as well as non-parametric settings \cite{gao2021efficient}.
On the other hand, it is important to understand the fundamental \textit{lower bounds} on the sample complexity needed by \textit{any} causal structure learning algorithm.
Such lower bounds have been established for the exponential family setting \cite{ghoshal2017information} and the linear equal-variance setting \cite{gao2022optimal}, but the lower bounds for a wide range of settings and assumptions remain uncharacterized.

Furthermore, since consistency of causal structure learning algorithms always requires some form of ``faithfulness" or genericity assumption (see \rref{sec:preliminaries}), there are likely tradeoffs between the strength of faithfulness assumption imposed and computational and statistical complexity.
Indeed, an interesting open question is to characterize the weakest assumption needed for causal structure learning, with the \textit{sparsest Markov representation} assumption \cite{raskutti2018learning} being one candidate.
Finally, the works discussed above are all in causally sufficient settings with only observational data.
Incorporating interventional data into these analyses would open the possibility for a reduction in overall sample complexity, and may introduce a landscape of tradeoffs between interventional and observational sample complexities.
Indeed, interventional data has been considered in recent works \cite{acharya2018learning,bhattacharyya2021efficient} on the statistical and computational complexity of \textit{causal inference} tasks, where the causal graph is assumed to be known and the task is to estimate interventional distributions.
An interesting future direction is to also explore the effect of interventional data on the complexity of causal structure learning.

\textbf{\textit{Experimental design for causal structure learning.}}
In this review article, we have focused on causal structure learning in a \textit{passive} setting, where we are given a dataset, or possibly several datasets from different interventions or contexts.
However, in many scientific settings, such as biology, where interventions such as genetic or chemical perturbations can readily be performed, an important component of causal discovery is the choice of what data to gather \cite{glocker2021causality}.
This leads us to consider \textit{experimental design} approaches for causal structure learning, where an experimenter may pick interventions (and their values) in an effort to identify the underlying causal structure.
Several approaches have been proposed for a variety of settings.
In the \textit{non-adaptive} setting, the experimenter picks all interventions at once.
In \cite{eberhardt2005number} it is shown that, in the absence of any pre-existing observational data, $p-1$ interventions are sufficient and in the worst-case necessary for identifying the underlying causal structure over $p$ variables.
Other work in the non-adaptive setting considers the presence of background knowledge (e.g., from observational data) \cite{hyttinen2013experiment}, differences in costs between interventions \cite{kocaoglu2017cost,lindgren2018experimental}, and a \textit{fixed-budget} setting \cite{ghassami2018budgeted}.

Alternatively, the \textit{adaptive} setting allows the experimenter to observe the outcome of each intervention before picking the next intervention.
\cite{he2008active} and \cite{hauser2014two} propose greedy approaches for the adaptive setting, picking new interventions based on some measure of either expected or worst-case information gain.
While these approaches are designed for the \textit{noiseless} setting, in which an infinite amount of data is gathered from each intervention, more recent works \cite{von2019optimal,tigas2022interventions} explore greedy approaches in the \textit{noisy} setting. 
\cite{greenewald2019sample} shows that strategies which maximize expected information gain can be exponentially sub-optimal in the number of interventions that they use, and propose the \textit{Central Node} algorithm for settings where the essential graph is a tree.
They show that this algorithm is a 2-approximation to the optimal adaptive strategy.
Follow-up work \cite{squires2020active} adapts this algorithm to a more general class of essential graphs, provides a characterization of the number of single-node interventions needed by an oracle to identify a causal graph, and shows that their algorithm uses within a logarithmic factor of this number of interventions.

In between the non-adaptive and adaptive settings, \cite{agrawal2019abcd} considers the active \textit{batched} setting, in which the experimenter observes the outcome of a batch of interventions before picking the next batch of interventions.
Recent work \cite{sussex2021near} establishes novel submodularity properties for greedy objectives in this settings, allowing for efficient optimization over the choice of interventions in each batch.
Taken together, these recent advances suggest several future directions, including (1) characterizing the number of multi-target interventions needed by an oracle in the adaptive case \cite{porwal2021almost}, (2) approximation guarantees for experimental design, compared to either oracles or optimal strategies, and (3) experimental design in settings with latent confounding \cite{kocaoglu2017experimental,addanki2020efficient}, selection bias, and cycles.

\textbf{\textit{Targeted causal structure learning.}}
Thus far, we have focused on the problem of causal structure learning as an end in itself; i.e., in both the passive and active settings discussed, the desired output was a causal graph (or equivalence class).
However, ultimately, a major motivation for causal structure learning is to use the causal model in downstream tasks.
A task of considerable importance is \textit{policy evaluation}, i.e., predicting the effect of an action.
The overall goal of task can be phrased as estimating a specific \textit{functional} of an interventional distribution defined by a structural causal model $M$.
Then two principal subtasks are (1) determining whether this functional is \textit{identifiable} by transforming it into a functional of the available distributions and (2) estimating the resulting functional from samples.
When the only available distribution is the observational distribution defined by $M$, possibly with some variables unobserved, the first subtask is covered by the \textit{ID algorithm} \cite{shpitser2006identification} and its variants \cite{shpitser2021proximal}.

More generally, data might be available from some set of interventional distributions defined by $M$, or from observational and interventional distributions associated to some related structural causal model $M'_1, \ldots, M'_K$.
The relation between these structural causal models is encoded using a \textit{selection diagram}, and the task of using the selection diagram to identify the functional is covered by a rich literature on \textit{transportability} \cite{bareinboim2012transportability,bareinboim2014transportability,lee2020generalized,correa2020general}.
Once the target functional is transformed into a functional of the available distributions, it becomes essential to estimate the functional in a sample-efficient way.
This has been extensively studied in the literature on \textit{semiparametric efficiency} \cite{rotnitzky2019efficient,bhattacharya2020semiparametric}, double machine learning \cite{jung2021estimating}, and targeted machine learning \cite{schuler2017targeted}, also covered in a recent review \cite{kennedy2022semiparametric}.
Thus far, causal structure learning and policy evaluation have been studied as separate tasks: the output of causal structure learning is a causal graph, while the input to policy evaluation is a causal graph or selection diagram.
Therefore, the current approach to using policy evaluation tasks when the graph is unknown would be to first perform causal structure learning, then to use the methods discussed for policy evaluation.
It is likely that this approach is not optimally sample-efficient - the two steps should be ``aware" of each other, i.e., causal structure learning should be performed in a way that is targeted toward the downstream task.

The problem of targeted causal structure learning remains mostly unexplored, with a few notable exceptions.
In the adaptive experimental design setting, \cite{agrawal2019abcd} considers targeted learning of any property of the underlying graph, and \cite{zhang2021matching} considers targeted learning of a ``matching" intervention, which affects the system in some desired way.
In the batched data setting, \cite{wang2018direct,zhao2019direct,Belyaeva_DCI} considers targeted learning of the \textit{difference} between two DAG models, instead of the DAG models themselves.
All of these works demonstrate computational and statistical benefits to targeted learning over untargeted structure learning, indicating that this is an important and promising direction.

\textbf{\textit{Causal structure in reinforcement learning.}}
Policy evaluation is also an important task in reinforcement learning, where the policy is a \textit{sequence} of actions that can depend on the state of the environment.
The overlap between reinforcement learning and causality has been recently explored in the simple setting of \textit{multi-armed bandits}, where an agent's actions do not affect the state of the environment.
By assuming that actions correspond to interventions in a known causal graph, the effects of different actions become related, allowing for better regret bounds \cite{lattimore2016causal,nair2021budgeted}.
If the causal graph is not assumed to be known, there is an additional exploration-exploitation tradeoff that needs to be taken into account, which has been considered in recent work \cite{de2020causal,lu2021causal,bilodeau2022adaptively}.
Since certain parts of the causal graph might not be relevant to predicting the effect of an action on some reward, the reinforcement learning setting is another case in which \textit{targeted} structure learning may be more efficient.

\section*{Acknowledgements}
Chandler Squires was partially supported by an NSF Graduate Fellowship. Caroline Uhler was partially supported by NSF (DMS-1651995), ONR (N00014-17-1-2147 and N00014-22-1-2116), the MIT-IBM Watson AI Lab, MIT J-Clinic for Machine Learning and Health, the Eric and Wendy Schmidt Center at the Broad Institute, and a Simons Investigator Award.

\bibliographystyle{spmpsci}      
\bibliography{references}

\end{document}

%% file: defs.tex
\usepackage{amsmath,amsfonts,amssymb}
\usepackage{bbm}
\usepackage{bm}
\usepackage{xcolor}

\newtheorem{example}{Example}
\newtheorem{thm}{Theorem}

\usepackage{prettyref}
\newcommand{\rref}[2][]{\prettyref{#2}}
\newrefformat{model}{Model\,\ref{#1}}
\newrefformat{listing}{Listing\,\ref{#1}}
\newrefformat{algm}{Algorithm\,\ref{#1}}
\newrefformat{line}{line\,\ref{#1}}
\newrefformat{sec}{Section\,\ref{#1}}
\newrefformat{subsec}{Section\,\ref{#1}}
\newrefformat{section}{Section\,\ref{#1}}
\newrefformat{appendix}{Appendix\,\ref{#1}}
\newrefformat{app}{Appendix\,\ref{#1}}
\newrefformat{def}{Definition\,\ref{#1}}
\newrefformat{defn}{Definition\,\ref{#1}}
\newrefformat{thm}{Theorem\,\ref{#1}}
\newrefformat{ax}{\ref{#1}}
\newrefformat{prop}{Proposition\,\ref{#1}}
\newrefformat{lemma}{Lemma\,\ref{#1}}
\newrefformat{cor}{Corollary\,\ref{#1}}
\newrefformat{corollary}{Corollary\,\ref{#1}}
\newrefformat{ex}{Example\,\ref{#1}}
\newrefformat{tab}{Table\,\ref{#1}}
\newrefformat{fig}{Fig.\,\ref{#1}}
\newrefformat{eqn}{Equation~(\ref{#1})}
\newrefformat{problem}{Problem\,\ref{#1}}
\newrefformat{assumption}{Assumption\,\ref{#1}}

\newcommand{\Cov}{\textrm{Cov}}
\newcommand{\Ber}{\mathsf{Ber}}
\newcommand{\Cat}{\mathsf{Cat}}

\newcommand{\indep}{\mathrel{\perp\mspace{-10mu}\perp}}
\newcommand{\kron}{\mathbbm{1}}

\newcommand{\cG}{\mathcal{G}}

\newcommand{\cI}{\mathcal{I}}
\newcommand{\cM}{\mathcal{M}}
\newcommand{\cN}{\mathcal{N}}

\newcommand{\bbE}{\mathbb{E}}

\newcommand{\bbD}{\mathbb{D}}
\newcommand{\bbP}{\mathbb{P}}
\newcommand{\bbM}{\mathbb{M}}
\newcommand{\bbX}{\mathbb{X}}
\newcommand{\bbZ}{\mathbb{Z}}

\DeclareMathOperator{\pa}{pa}

\DeclareMathOperator{\spo}{sp}

\DeclareMathOperator{\skel}{skel}
\DeclareMathOperator{\vstruct}{vstruct}

\DeclareMathOperator{\discr}{discr}
\DeclareMathOperator{\indepmodel}{\cI_{\indep}}

\DeclareMathOperator{\pre}{pre}

\newcommand{\parspace}{\vspace{0.4em}}